\documentclass[twocolumn]{aa}
\usepackage{natbib}
\usepackage{graphicx}
\usepackage{psfig}
\usepackage{lscape}
\usepackage{rotating}
\usepackage{longtable}
\input epsf
\newcommand{\ha}{$\mbox{H}\alpha \ $}
\newcommand{\He}{\mbox{He}\,{\sc ii} \,}
\newcommand{\he}{$m_{\lambda 4684} \ $}
\newcommand{\offhe}{$m_{\lambda 4781} \ $}

\newcommand{\N}{N~{\sc iii}\,}
\newcommand{\BB}{broad-band B\ } 

\newcommand{\Nfive}{N\,{\sc v}\,}
\newcommand{\Nthree}{N\,{\sc iii}\,}
\newcommand{\Cthree}{C\,{\sc {iii}}\,} 
\newcommand{\Cfour}{C\,{\sc iv}\, }

\newcommand{\ciii}{$\lambda 4650$\, }
\newcommand{\heii}{$\lambda 4686$\, }
\newcommand{\redciii}{$\lambda 5696$\, }
\newcommand{\civ}{$\lambda 5812$\, }
\newcommand{\blue}{\Cthree$\lambda$4650/\He$\lambda$4686}
\newcommand{\extended}{$^{c}$}
\newcommand{\red}{$^{b}$}
\newcommand{\factor}{$^{a}$}
\begin{document}
\title{A Spectroscopic Search for the non-nuclear Wolf-Rayet Population 
of the metal-rich spiral galaxy M\,83
\thanks{Based on observations made with ESO Telescopes at the Paranal Observatory under programme ID 69.B-0125}}
\titlerunning{WR population of M\,83}
	\author{ L. J. Hadfield\inst{1}
 	\and Paul A. Crowther\inst{1} \and H. Schild\inst{2} \and W. Schmutz\inst{3}}
	\offprints{L. J. Hadfield}
         \institute{Department of Physics \& Astronomy, University of Sheffield, Hicks Building, Hounsfield Rd,
   Sheffield, S3 7RH, United Kingdom\\
            \email L.Hadfield@sheffield.ac.uk
        \and  Institut f\"ur Astronomie, ETH-Zentrum, CH 8092 Z\"urich, 
                Switzerland        \and
          Physikalisch-Meteorologisches Observatorium, CH-7260 Davos,  Switzerland}
	\abstract{We present a catalogue of non-nuclear regions
	  containing Wolf-Rayet stars in the metal-rich spiral galaxy
	  M\,83 (NGC\,5236).  From a total of 283 candidate regions
	  identified using He\,{\sc ii} $\lambda$4686 imaging with
	  VLT-FORS2, Multi Object Spectroscopy of 198 regions was
	  carried out, confirming 132 WR sources.  From this
	  sub-sample, an exceptional content of $\sim$1035$\pm$300
	  WR stars is inferred, with N(WC)/N(WN)$\sim$1.2, continuing
	  the trend to larger values at higher metallicity amongst
	  Local Group galaxies, and greatly exceeding current
	  evolutionary predictions at high metallicity.  Late-type
	  stars dominate the WC population of M\,83, with
	  N(WC8--9)/N(WC4--7)=9 and WO subtypes absent, consistent
	  with metallicity dependent WC winds. Equal numbers of late
	  to early WN stars are observed, again in contrast to current
	  evolutionary predictions. Several sources contain large
	  numbers of WR stars. In particular, \#74 (alias region 35
	  from \citeauthor{devau83}) contains $\sim$230 WR stars, and
	  is identified as a Super Star Cluster from inspection of
	  archival HST/ACS images. Omitting this starburst cluster
	  would result in revised statistics of N(WC)/N(WN)$\sim$1 and
	  N(WC8--9)/N(WC4--7)$\sim$6 for the `quiescent' disk
	  population. Including recent results for the nucleus and
	  accounting for incompleteness in our spectroscopic sample,
	  we suspect the total WR population of M\,83 may exceed 3000
	  stars.
}
   \maketitle
%
\section{Introduction}

Massive stars play a major role in the ecology of galaxies via radiative,
mechanical and chemical feedback \citep{smith05}. Wolf-Rayet (WR)
stars in particular, albeit rare and short-lived, make a significant
contribution  to their environment
via the mechanical return of nuclear processed material to the
interstellar medium (ISM)  through 
their  exceptionally powerful stellar winds.

Metallicity, $Z$, is a key factor in determining the  number and
subtype distribution of a WR
population. Although the metallicity dependence of WR wind properties
still remains unclear, mass-loss prior to this phase has been
established to depend on metallicity, with the latest models
predicting \.{M} $\propto Z^{\sim 0.8}$  for O stars \citep{vink01}.  Evolutionary
models for single stars predict that the minimum mass cut-off required
for WR formation should decrease as metal content increases.  It is anticipated that
the minimum mass required for progression to the WR phase decreases
from $\sim 32 \mbox{M}_{\,\odot}$ for a SMC-like metallicity to $\sim 21
\mbox{M}_{\,\odot}$ for super-Solar metallicity \citep{meynet04}.  
Single star predictions are broadly consistent with the initial masses of
WR stars in the Milky Way, LMC and SMC inferred from cluster membership
\citep{massey00, massey01}. The fractional distribution of carbon-rich (WC)
 to nitrogen-rich (WN) stars  is also known to increase with metallicity, such 
that empirically one would expect to observe a large WC population in a metal-rich 
environment \citep{massey98}. The formation of WR stars at low metallicity 
is anticipated primarily via close binary Roche Lobe 
Overflow (RLOF), yet the observed
WR binary fraction in the SMC does 
not differ from that of the Milky Way \citep{foellmi03}.

With the aim of substantiating such predictions, surveys of WR stars
in Local Group galaxies (typically 0.2--1Z$_{\odot}$) have been
carried over the last two decades.  At sub-Solar metallicities, the
LMC and SMC have been well sampled \citep{breysacher99, massey03}, as
has M\,33 \citep{massey98, abbott04}. In contrast, M\,31 is the only Local
Group member with super-Solar metallicity, but its unfavourable
inclination and large spatial extent makes surveying the complete WR
population very challenging.  In order to increase the variety of
galaxies sampled, our group has begun to look beyond the Local Group
\citep[e.g. NGC300,][]{schild03}. 

Galaxies hosting substantive WR populations are known as `WR galaxies'
\citep{kunth81, scp99}, where the number of WR stars ranges from
$\sim$35 in NGC1569-A \citep{gonzalez97} to 2$\times 10^{4}$ in
Mrk 309 \citep{schaerer00}. Within specifically metal-rich environments, 
previous studies of WR populations have generally been restricted to 
integrated spectra from bright star forming knots 
\citep[e.g.][]{schaerer99} or H\,{\sc ii} regions \citep[e.g.][]{pindao02}.
Here we present the results of a deep imaging and spectroscopic survey of the 
disk WR population within the metal-rich galaxy M\,83, in which WR signatures
have previously been identified by \cite{rosa87} and \cite{bresolin02}.

M\,83 (NGC 5236) is a massive, grand-design southern spiral (SBc(s)\,II)
with on-going star formation in its spiral arms plus an active
nuclear starburst \citep{elme98, harris01}.  M\,83
is the principal member of a small galaxy group ($\sim
11$ members) within the Centaurus A complex \citep{kara02}. 
Located at a distance of 4.5$\pm$0.3Mpc
  \citep{thim03}, its favourable inclination and apparently 
high metal abundance of
log(O/H)+12=9.2 \citep{bresolin02} makes M\,83 an ideal candidate for
studies of massive stellar populations at high metallicity.  

More
recently, oxygen abundances in metal-rich galaxies have been revised
downward \citep{pilyugin04, bresolin04}, such that M\,83 may have a
metal abundance closer to log(O/H)+12=9.0 (Bresolin, 2004,
priv. comm.), i.e. approximately twice the Solar oxygen abundance
of log(O/H)+12=8.66 recently derived by \citet{asplund04}.


We present the results of an imaging and spectroscopic survey of the
WR content of M\,83 using the ESO Very Large Telescope (VLT). The
present paper complements the initial findings of this study reported
in \citet[hereafter Paper\,I]{pac04}.
In Sect.\,\ref{obs:red} we briefly describe the observations and data
reduction techniques employed. Section \ref{wr:con} discusses the method
followed to obtain a global WR population of M\,83.
Sect.\,\ref{discussion} discusses the properties of metal rich WR stars
with those of Local Group galaxies and evolutionary models. 
Finally, conclusions are drawn in Sect.\,\ref{conc}.

\section{Observations and data reduction}
\label{obs:red}
We have observed M\,83 with the ESO Very Large Telescope UT4 (Yepun)
and Focal Reduced/Low Dispersion Spectrograph \#2 (FORS2).  The
detector consists  of a mosaic of two $2048 \times 1024$ MIT/LL CCDs which
in conjunction with the standard collimator provides a field-of-view
$6.8'\times6.8'$ and an image scale of 0.126$''$/pixel. Photometric
observations of M\,83 were made between May--June 2002 with follow-up
spectroscopic data being acquired during April--June 2003.

\subsection{Imaging}
\label{imaging}

\begin{figure}
\includegraphics[width=9cm]{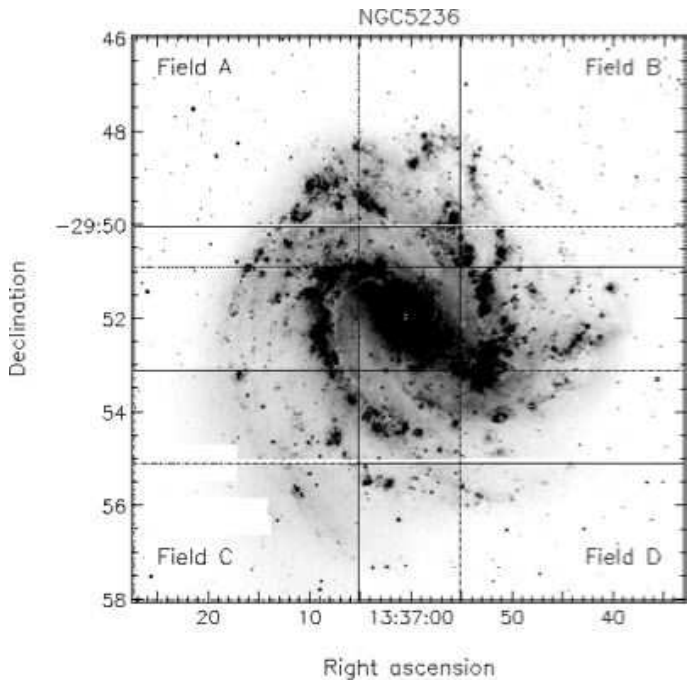}
\caption{Four combined \ha VLT FORS2 frames of M\,83.  The overall size of the image is  $\sim 12\,'
 \times 12\,'$.  The fields A, B, C and D used to image M\,83 are
 marked.  North is up and east is to the left of the image.}
\label{fields}
\end{figure}

M\,83 subtends 12.9$'$ by 11.5$'$ on the sky, preventing it being
imaged by a single FORS2 frame.  In order to obtain complete coverage,
the galaxy was divided into four overlapping regions, covering the
NE (Field A), NW (B), SE (C) and SW (D) as indicated in 
Fig.\,\ref{fields}.  Occulting bars were positioned in Field C to
prevent detector saturation by bright foreground stars.  The
central 15$''$ appears saturated on all images obtained, and as a
result the WR population of the nucleus can not be discussed further.

FORS2 was used on 2 June 2002 to obtain narrow-band images with
central wavelengths 4684\AA , 4781\AA \, and band widths of 66\AA \,
and 68\AA \, respectively.  These were obtained consecutively for each
Field in seeing conditions between 0.6 -- 0.8$''$\footnote{0.8$''$
corresponds to a linear scale of 18pc at the distance of M\,83
\citep{thim03}} with individual exposures of 1800s.  The $\lambda$4684
filter is coincident with the strong WR emission features which
incorporates the N\,{\sc iii} ($\lambda 4640$\AA), \Cthree ($\lambda
4650$\AA) and \He ($\lambda 4686$\AA) emission lines, whereas the
latter samples a wavelength region relatively free from emission,
providing a measure of the continuum level. In addition to these, 2
exposures (60s and 600s) were taken using narrow-band on- and off-\ha
\,filters $(\lambda 6563,\,6665\,\mbox{\AA}, \,
\mbox{FWHM}=61,\,65\,\mbox{\AA}$) on 16 May 2002. Finally, in order to
supplement the primary dataset, 2 exposures (60s and 120s) were also
acquired using a Bessell B filter on 21 May 2002.

%
%
\subsection{Photometry}

Images were prepared following standard procedures i.e. debiased, flat
field corrected and cosmic ray cleaned.  Photometry of individual
sources within M\,83 was performed using the package {\sc daophot}, a
point-spread function (PSF) fitting routine within {\sc iraf}. 
%
%
Absolute photometry in the \BB filter was achieved with the aid of
photometric standard fields Ru 152 and PG 1528+062 (containing a total
of 10 photometric standards, 11.9 $\leq$ B $\leq$16.3).  For the
narrow-band images such standards are not available and photometric
zero-points have been obtained by observing spectrophotometric
standards LTT7987 (B = 12.2) and G138-31 (B = 16.5).

 The majority of our sources appear point-like on the ground-based
images. However, a number of bright sources are surrounded by a faint,
extended halo, which was not accounted for in the PSF photometry and
as a result only a lower limit to the magnitude is given, based on PSF
photometry.  A further subset of the bright sources are spatially
extended, indicating that PSF photometry is inappropriate, as
indicated in Table~A1 in the appendix.

Typical formal photometric errors range between 0.02\,mag
($\sim$18\,mag), 0.05\,mag ($\sim$20\,mag) and 0.08\,mag
($\sim$22\,mag). Significantly higher errors, of up to 0.15\,mag, are
obtained for regions of the galaxy where the background levels are
high, or they are located in spatially crowded regions.

As a consistency check we have compared results obtained for the two
Bessell B exposures (for which the PSF model was based on different
template stars) and also derived magnitudes for objects which 
appear in multiple fields.  Excellent agreement was observed in
both cases, with results agreeing to within the formal errors.  In a
minority of cases this was not achieved due to severe
crowding.


\subsection{Candidate Selection}

WR candidates were identified by searching for He\,{\sc ii} / \Cthree\, 
excess emission (at $\lambda$4684)  relative to the continuum
($\lambda$4781), i.e. a negative  value of $\Delta m$ =  \he - $m_{\lambda 4781}$.
 The optimal method of identifying suitable candidates 
was found to be via `blinking' individual $\lambda$4781 and $\lambda$4684 frames together with the
 difference  image obtained by subtracting the $\lambda$4781 image from
 the $\lambda$4684 frame. In total, 283 candidate $\lambda$4684 emission sources were identified.

For 75\% of our candidates we have obtained a magnitude in at least
the $\lambda$4684 filter.  For cases where we did not obtain photometry, the
object was either too faint or was located in a spatially crowded
region. In addition, for a significant fraction of the fainter sources
  it was not possible to measure a $\lambda$4871 magnitude.

\begin{figure}
\includegraphics[width=7.8cm,angle=0]{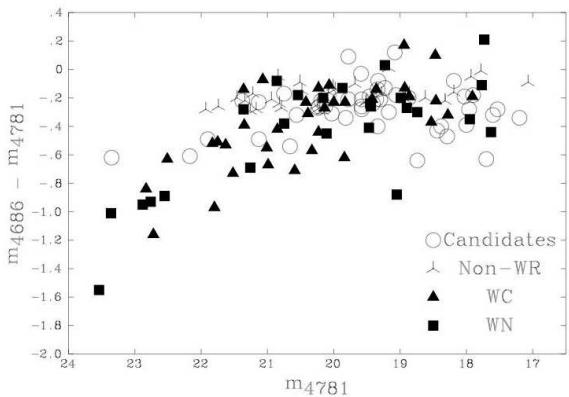}
 \caption{Comparison between \offhe \ magnitude and \he excess  of the WR
 candidates located in M\,83.  Regions which have been
 spectroscopically observed and subsequently eliminated or classified
 as WR regions are presented in the key.  Regions which still await
 spectroscopic observations are also marked. Sources for which PSF
 photometry was unavailable are not marked.}
 \label{photom:excess}
\end{figure}

Candidates were grouped according to continuum brightness, $\Delta m$,
and association with underlying H\,{\sc ii} regions.  To ensure we
spectroscopically observed a representative sample, a selection from
each group was chosen for spectroscopic follow up.  In
Fig. \ref{photom:excess} we show $\Delta m$ as a function of continuum
magnitude for the sources in which WR signatures were either
spectroscopically confirmed, rejected or no spectroscopy was obtained,
i.e. the remaining candidates. The majority of confirmed sources have
a $\lambda$4684 excess between --1.5 $\leq \Delta m \leq$ --0.4 mag,
although a few do exhibit rather smaller values of $\Delta
m$. In contrast, all rejected regions  have $\Delta m \geq$ --0.2 mag,
suggesting that remaining candidates which display a moderate
$\lambda$4684 excess should represent regions that genuinely host WR
stars, together with a subset of those for with $\Delta m \sim$ 0.0
mag.

\subsection{Spectroscopy}

Spectroscopic data was obtained using FORS2 with the Multi Object
Spectroscopy (MOS) mode.  MOS datasets of  individual WR candidates were
obtained during seeing conditions of $\sim$0.5--1.0$''$, using a slit 
width of $0.8 ''$.  The CCD was binned by a factor of 2
in the dispersion direction, resulting in a dispersion of
3.3\AA\,pixel$^{-1}$ with the 300V grism
and a spectral resolution of $\sim 7$\AA, as
measured from comparison arc lines. The wavelength range of individual targets
depended on their position within the MOS mask but typical wavelength
coverage was $\sim$3700\AA\ to $\sim$7500\AA\ .  

MOS  allows the spectra of up to 19 candidates
to be recorded simultaneously. However, due to positional limitations
this was generally restricted to $\sim$15, supplemented where
possible by H\,{\sc ii} regions.  In total, 198 candidates have 
been spectroscopically observed using 17 different MOS masks.  To maximise
continuum S/N, sources were grouped according to brightness, with
total on-source integration times ranging from 720s for the brightest
objects to 4800s for the faintest.  Details of the spectroscopic log can be found in 
Table\,\ref{Tab:spec}, and includes DIMM seeing measurements.
The MOS masks were labelled according to the region of M\,83 in which
they were observed, i.e. Field A was observed using 5 different 
masks labelled A1 to A5. 

\begin{table}
\begin{center}
\label{Tab:spec}
\caption[]{FORS2 Multi Object Spectroscopy (MOS) observing log for M\,83.}
\label{Tab:spec}
\begin{tabular}{lclc}
\hline
Date &MOS &Exposure &DIMM Seeing\\ 
&Mask& \multicolumn{1}{c}{(sec)}& ($''$)\\
\hline
2003-04-06&D2& $3 \times 900$& 0.6\\
\hline
2003-04-07&D1&  $1 \times 2400$& 0.5\\
&D4&$3 \times 240$& 0.5\\
\hline
2003-04-13&D1& $1 \times 2400$& 0.7\\
&D3&  $3 \times 500$& 0.7\\
&D5& $3 \times 600$& 0.5\\
\hline
2003-05-20&C1&$2 \times 2400$& 1.0\\
\hline
2003-05-21&C2&$3 \times 900$& 0.9\\
&C3&$3 \times 240$& 0.9\\
\hline
2003-05-24&A1&$ 1 \times2400$& 0.6\\
&A3&$ 3 \times 500$& 0.5\\
\hline
2003-05-26&A2,B2&$3 \times 900$& 0.7, 0.8\\
&A5&$ 3 \times 600$& 0.8\\
&A4&$ 3 \times 240$& 0.6\\
\hline
2003-06-17&A1&$1 \times 2400$& 0.6\\
&B1&$2 \times 2400$& 0.6, 0.5\\
&B3&$3 \times 250$& 0.7\\
&B4&$3 \times 240$& 0.5\\
\hline\hline
\end{tabular}
\end{center}
\end{table}

Datasets were prepared and processed using standard {\sc iraf} and
{\sc figaro}  packages i.e. the data were bias subtracted, flat
field corrected, spectra were traced, extracted, and subsequently 
wavelength and flux calibrated. For very  faint sources, where no 
continuum  was  evident, identification was made solely on the 
basis of strong emission lines, a neighbouring  continuum source was
used as the trace.

Spectroscopic magnitudes (hereafter m$_{\rm spec}$) were obtained by
convolving the individual spectra with  the 
transmission curves of the imaging filters.  A comparison between the 
spectroscopic and the PSF photometry at $\lambda 4684$  permitted absolute 
flux calibration.  For 160 sources brighter than 24 mag the average slit 
correction factor was found to be 3.1 ($\sigma=1.9$).  This large factor 
is due in part to the often crowded nature of sources, such that the full 
profile extent was not extracted.  For 20 spectroscopically observed 
regions, where photometry was unavailable, we corrected the spectroscopy by
a factor of 3$\pm$1.5.

The blue ($\sim$4500\AA) continuum S/N ranged from $\sim$80 for 
the brightest sources, to $\leq$1 for the faintest sources. 
Nevertheless, lines were typically detected even in the faintest sources 
at the 5--10$\sigma$ level, and a source was only confirmed if WR emission
lines were detected at a $\geq 3\sigma$ level. 

\section{The Wolf-Rayet Population of M\,83}
\label{wr:con}
The WR content of the disk of M\,83 has been determined by visually
inspecting the extracted spectra for the characteristic WR emission
signatures, i.e.  \Nthree$\lambda$4634--41, \Nfive$\lambda$4603--20,
\Cthree \ciii -- \He \heii blend (hereafter blue WR features) and/or
C\,{\sc iii} $\lambda$5696, \Cfour \civ features (hereafter yellow WR
features).

Of the 198 sources spectroscopically  observed, 132 contain WR
emission features at a 3$\sigma$ level.  These are presented in a
catalogue in Table A1, in the Appendix, which includes coordinates,
PSF (or spectroscopic) photometry, interstellar reddening, line measurements and  
WR populations. Deprojected distances are included using M\,83 
properties presented in Table\,1 of \citet{lundgren04}. 
We present finding charts for all confirmed sources
in the (electronic only) Appendix, Figures B1--B17.

\begin{figure}[htbp]
\begin{center}
\includegraphics[width=6cm,angle=0]{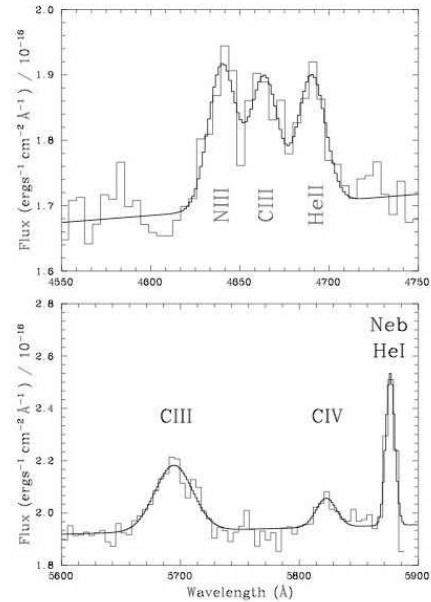}
\end{center}
\caption{ Observed, velocity corrected ($v_r = 513 \mbox{km\,s}^{-1}$)
  blue (top) and yellow (bottom) spectral regions of  Source
  \#74. Gaussian line profiles are overplotted for the WR
  features: \Nthree ($\lambda 4634-41$), \Cthree
  ($\lambda 4647-50$) and \He ($\lambda 4686$) in the blue,
 and \Cthree ($\lambda 5696$) and \Cfour ($\lambda 5801-12$)
 in the yellow.}
\label{fig:fits}
\end{figure}

Of the remaining 66 sources, 40 displayed an early-type spectrum
with no WR emission present, 16 resembled that of a late-type carbon
star, while 10 sources revealed WR features below the 3$\sigma$ level, or
lacked the blue region necessary for WN identification.
The latter two groups, along with the 79 regions which were not
spectroscopically observed, are given in our candidate list
(Table\,A2 in the Appendix).

\subsection{Interstellar reddening}

 Estimates of the interstellar reddening for our confirmed WR sources have
generally been derived using measurements of the nebular H$\alpha$
(accounting for nearby [N\,{\sc ii}] emission) and
H$\beta$ features present in the extracted spectra. 

Assuming Case B recombination theory for typical electron densities of
$10^{2}\,\mbox{cm}^{-3}$ and a temperature of $10^{4} \mbox{K}$
\citep{hummer87}, we obtain  0.2$\leq$E(B-V)={\emph{c}}(H$\beta$)/1.46
$\leq$ 0.8 mag for the majority of the sources, with a few outliers,
and typical formal uncertainty of $\pm$0.02 mag.  Where 
Balmer emission was observed,  typical H$\beta$ equivalent widths lay in 
the range $\mbox{and} \sim 20 \ \mbox{to} \sim 150$\AA.  Consequently, the 
underlying stellar absorption components ($\leq$1\AA\ at H$\beta$) 
are neglected. 

In 41 sources no nebular lines were observed. For those with a
well defined continuum, E(B--V) was estimated by assuming an 
intrinsic optical flux distribution equivalent to a late O-type star,
with typical uncertainty of $\pm$0.05--0.1 mag. In  15 cases, the continuum 
S/N was insufficient for this comparison and an average 
reddening of E(B--V)= 0.5$\pm$0.3 was adopted. Correction for
reddening adopt a standard \citet{seaton79} extinction law with
R=3.1=$A_{\rm V}$/E(B-V).

\subsection{Spectral classification}

In order to classify and quantify the WR population within each
region, we have fit Gaussian line profiles to the blue and yellow WR
features, revealing line fluxes, equivalent widths and FWHM.  An
example of the fits to the blue and yellow WR features is presented
in Fig.\,\ref{fig:fits}, where a source (\#74) hosting a mixed WN and
WC population is presented.

In general, it was straightforward to distinguish between WN (strong
\He $\lambda 4686$) and WC subtypes (strong \Cthree $\lambda$4650 and
\Cthree $\lambda$5696 and/or \Cfour $\lambda$ 5801-12).  The following
classification scheme was applied for further subdivision. In a
minority of cases it was not possible to separate the $\lambda$4650 --
$\lambda$4686 features into individual components, and as a result an
overall blend was measured.  Since WC subtypes were assigned on the
basis of \redciii and \civ features, this did not prevent accurate
classification.

Late and early  WN subtypes were assigned if \He $\lambda 4686$ emission
was accompanied by \N $\lambda 4634-41$ or \Nfive $\lambda 4603-20$ emission,
respectively. If nitrogen lines were undetected, we assigned a  WNE subtype
if  FWHM (\He $\lambda 4686$) $>$ 20\AA, and WNL otherwise. For WC stars,
 WC4 -- 6 was assigned if \Cfour $\lambda 5801-12$ was present
along with either weak or absent \Cthree $\lambda 5696$.  For $0.25
\leq F_{\lambda}$ (\Cthree $\lambda$5696/\Cfour $\lambda$5801-12) 
$\leq 0.8$ sources were classified WC7, and WC8--9 if  \Cthree $\lambda 5696$
was present, with \Cfour $\lambda 5801-12$ weak or absent. 

\begin{figure}[htbp]
\begin{tabular}{c}
\includegraphics[width=7cm,angle=0]{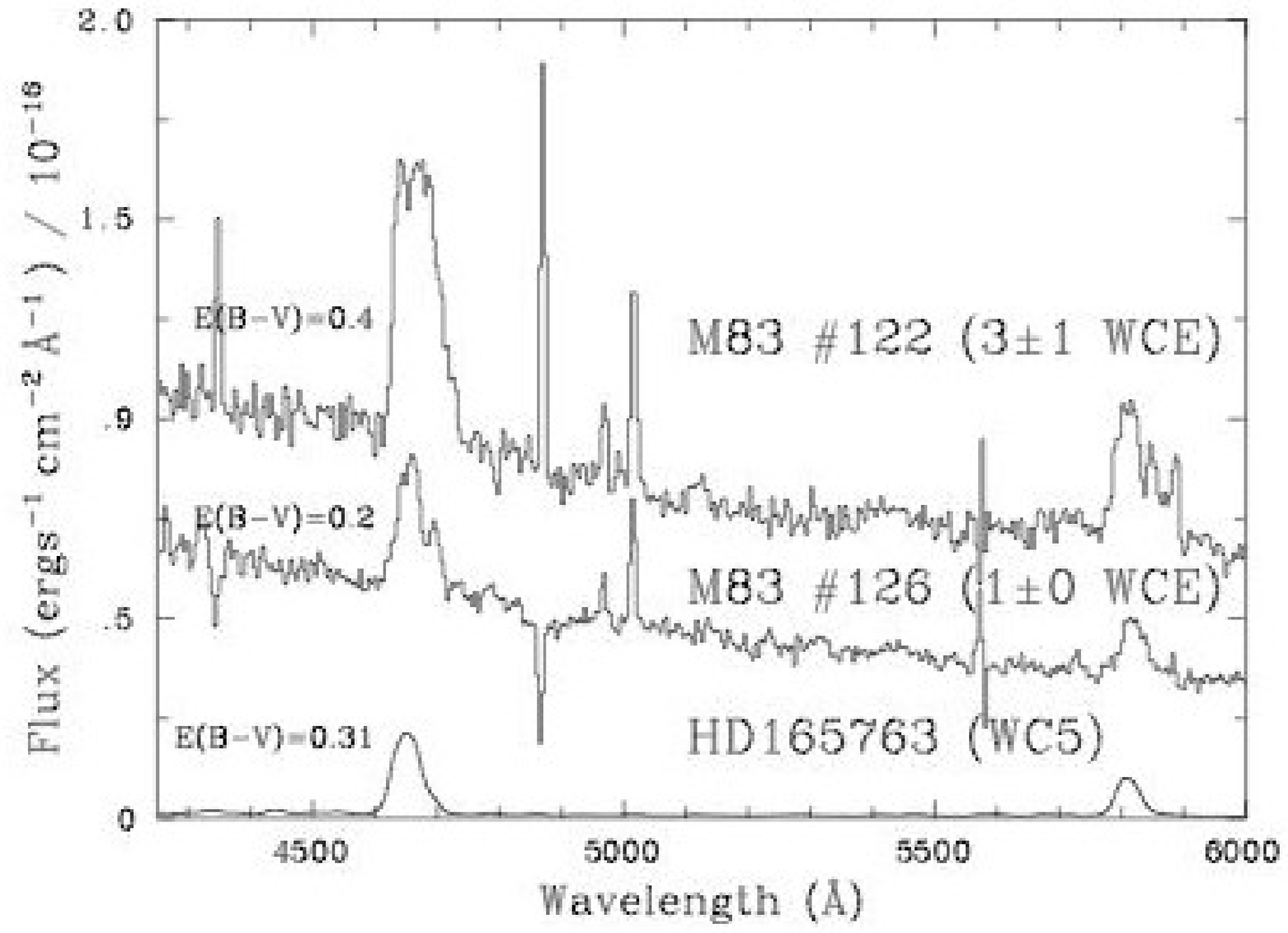}\\
\includegraphics[width=7cm,angle=0]{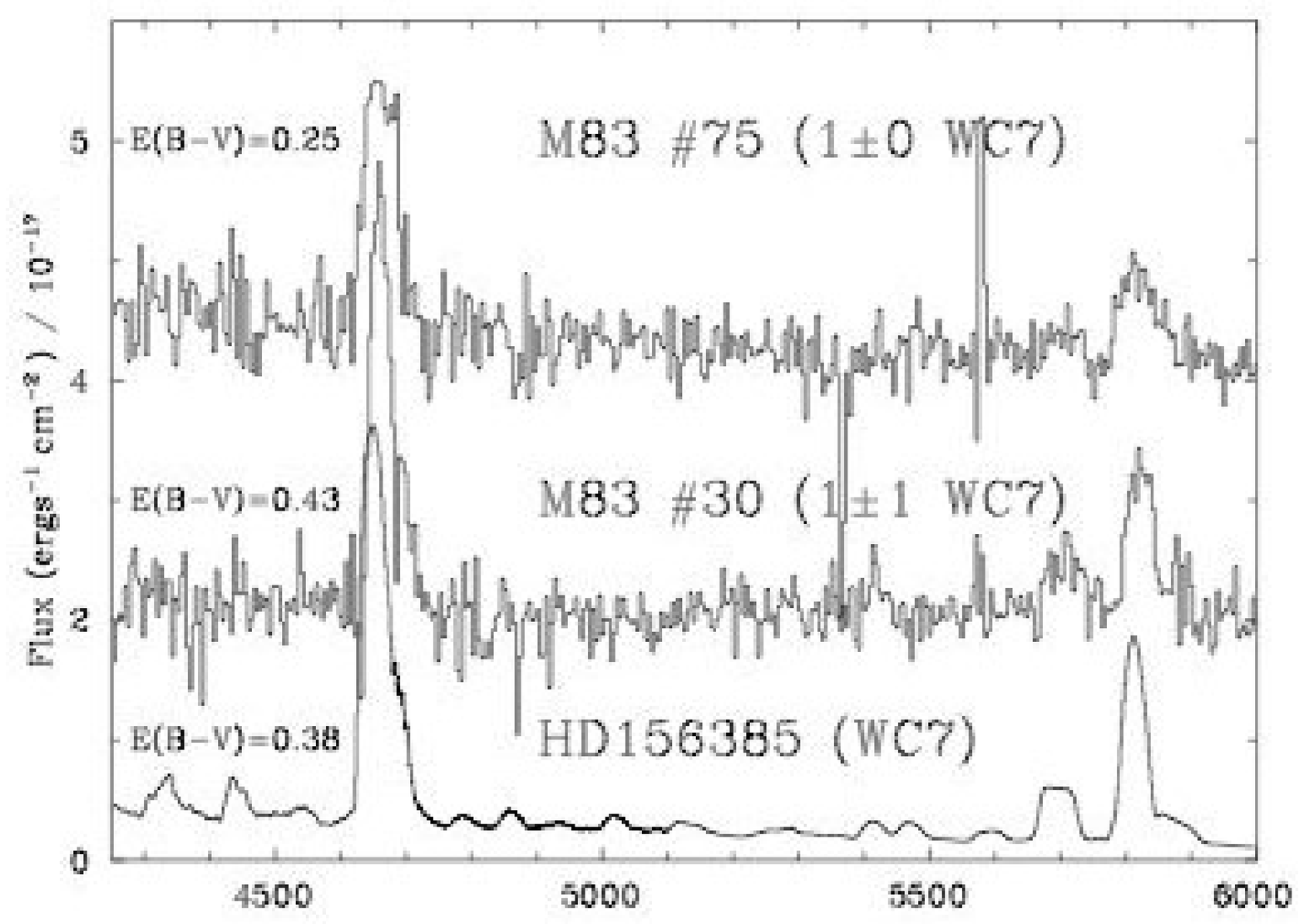}\\
\includegraphics[width=7cm,angle=0]{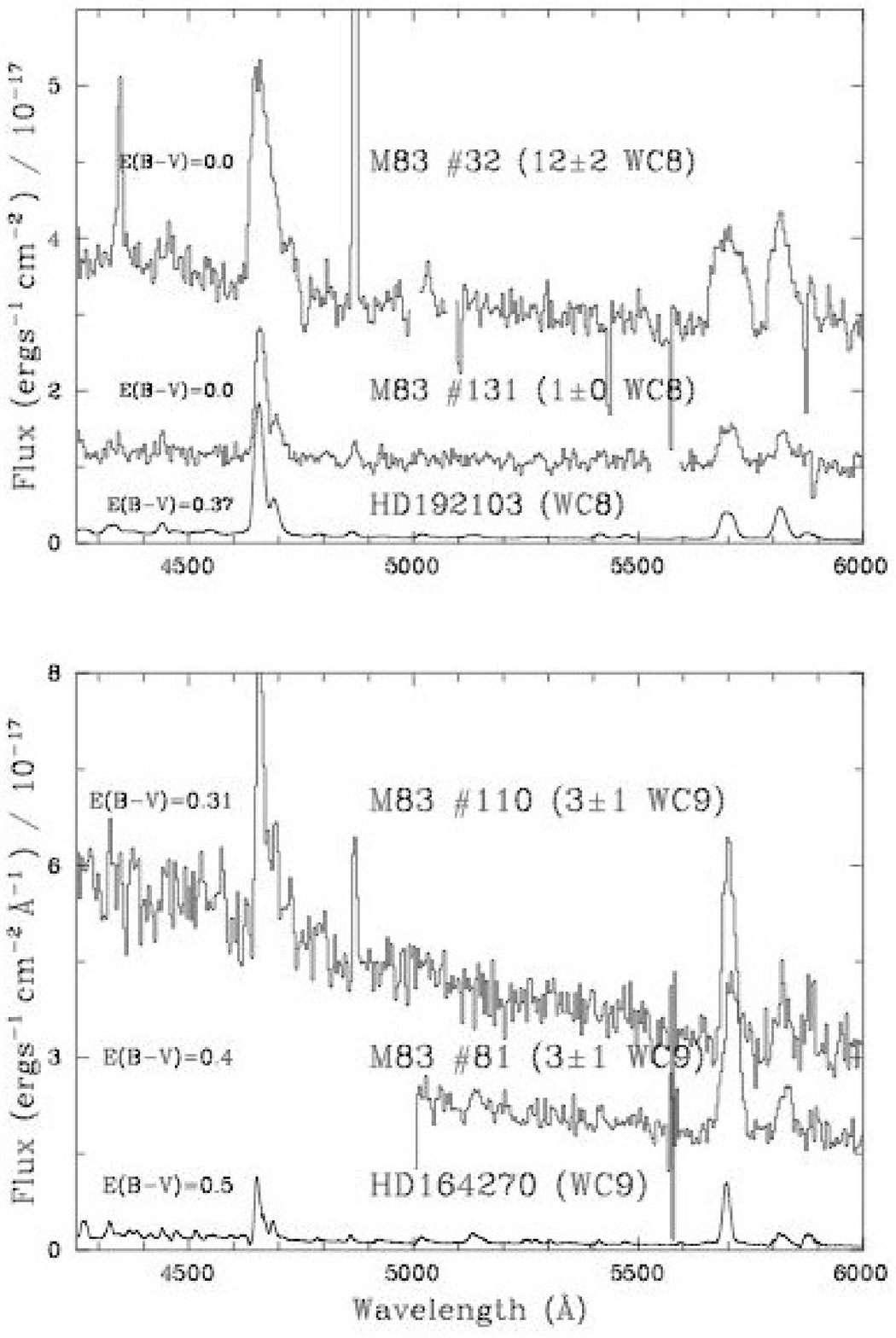}
\end{tabular}
\caption{ De-reddened spectral 
comparison between WC members in M\,83 with Galactic WC stars scaled to
the distance of M\,83 (Galactic distances from \citet{derhucht01}).  To
avoid confusion, WCE and WC7 sources are offset by
2$\times 10^{-17}$ erg\,s$^{-1}$\,cm$^{-2}$\AA$^{-1}$ whereas WC8 and WC9
spectra are offset by 1$\times 10^{-17}$
erg\,s$^{-1}$\,cm$^{-2}$\AA$^{-1}$.} 
\label{spectra}
\end{figure}

\begin{figure}[htbp]
\centering
\includegraphics[width=7cm,angle=0]{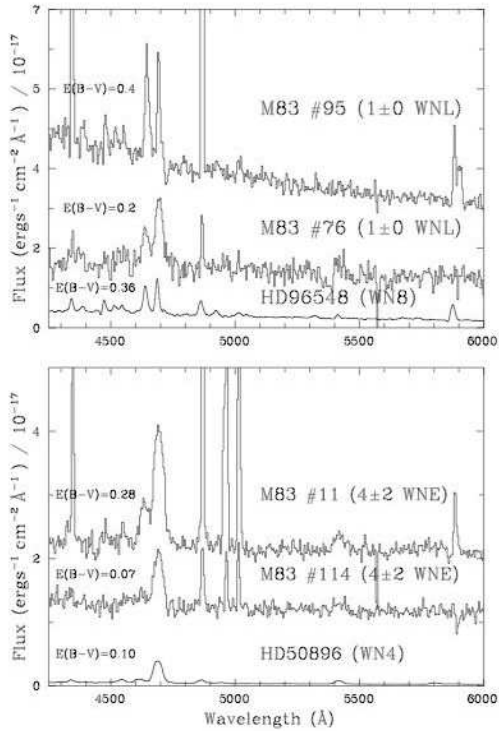}
 \caption{De-reddened spectral 
comparison between WN  complexes in M\,83 with individual Milky Way
WN stars scaled to the distance of M\,83 (Galactic distances from \citet{derhucht01}).
To avoid confusion, individual sources are successively offset by 1$\times 10^{-17}$ 
erg\,s$^{-1}$\,cm$^{-2}$\AA$^{-1}$.}
\label{WN_sp}
\end{figure}

To ensure consistency with previous studies \citep[e.g.][]{schaerer99,
bresolin02, chandar04} we have derived WR populations based on
individual line fluxes adapted from \citet*{schaerer}. As discussed in
Paper\,I, we adopt He\,{\sc ii} $\lambda$4686 lines fluxes of
5.2$\times 10^{35}$ erg\,s$^{-1}$ and 1.6$\times 10^{36}$
erg\,s$^{-1}$ for WN2-5 and WN6-10 stars, respectively.  For WC stars,
we adopt C\,{\sc iv} $\lambda$5801 line fluxes of 1.6$\times 10^{36}$
erg\,s$^{-1}$ and 1.4$\times 10^{36}$ erg\,s$^{-1}$ for WC4--6 and WC7
stars, respectively, and a C\,{\sc iii} $\lambda$5696 line flux of
7.1$\times 10^{35}$ erg\,s$^{-1}$ for WC8--9 stars. WR contents of
individual sources then follow, with populations rounded to the
nearest integer ($\geq$1). In one source (\#117), we were unable to
reliably extract the spectrum since it was located at the very edge of
the slit, and so a measure of the reddening/line flux was not
possible. Nevertheless, broad He\,{\sc ii} $\lambda$4686 is clearly
present, with no WC signature, such that we indicate a population of
$\geq 1$ early-type WN star.

In Fig.\,\ref{spectra} we compare sources containing representative
late, mid and early WC stars from M\,83 with extinction corrected Milky
Way counterparts, scaled to the distance of M\,83. Large line widths amongst
M\,83 members hosting late WC stars are apparent, particularly for \#32
versus HD\,192103 (WC8) and \#81 versus HD\,164270 (WC9). In contrast,
sources containing WC4--7 stars indicate similar line widths to
individual Galactic counterparts. 
Comparisons between sources containing WNL and WNE stars in M\,83 and
two Galactic counterparts are shown in Fig.\,\ref{WN_sp}, revealing similar
spectral morphologies. Other examples of sources hosting WN and WC populations
are presented in Fig.1 of Paper\,I.

\subsection{The M\,83 WR population -- individual stars, binaries,
 complexes or clusters?}

\begin{figure*}[t!]
\begin{tabular}{cc}
\includegraphics[width=7.8cm,angle=0]{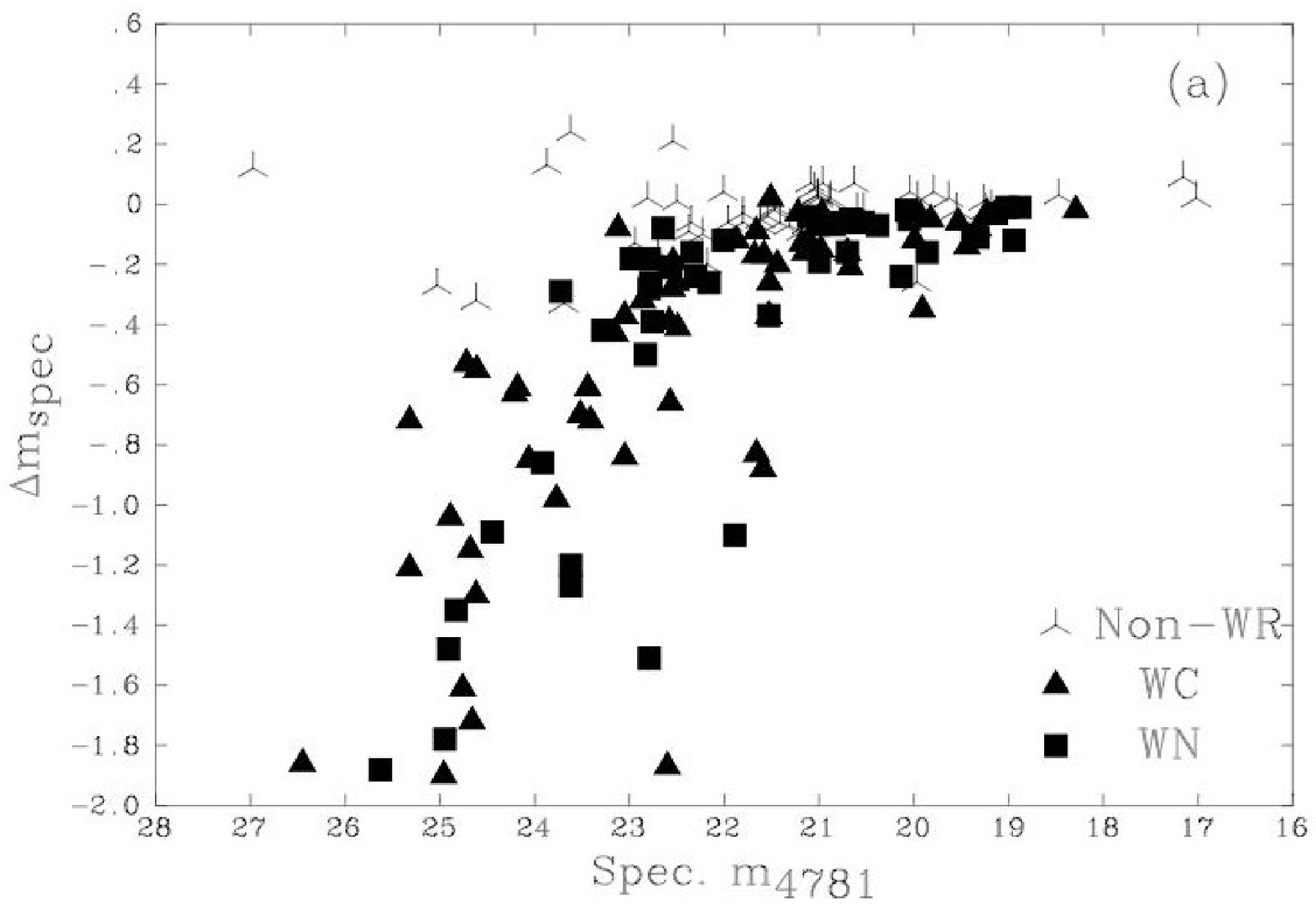}
\includegraphics[width=7.8cm,angle=0]{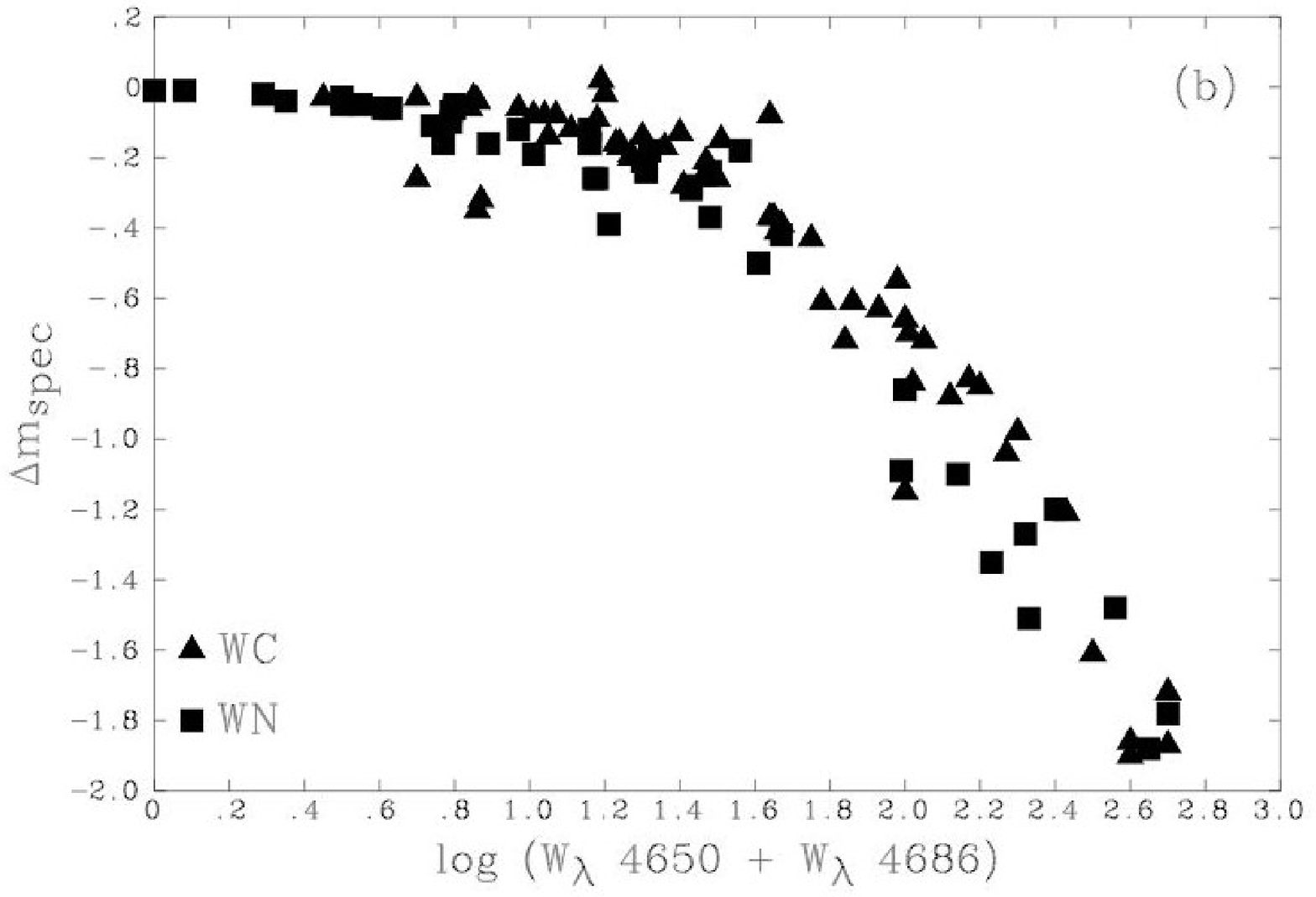}
\end{tabular}
\caption{Comparison between (a) the spectroscopic \offhe  magnitude
and (b) line equivalent width of the \blue\ WR features to the
spectroscopic \he excess.  Two WC objects, \#96 and 124, are not
marked, since their spectra start longward of He\,{\sc ii}. The left
panel confirms that regions without WR signatures are clustered around
$\Delta m_{\rm spec} \sim$0$\pm$0.2 mag, whilst regions with confirmed
WR signatures span a wide range, up to $\Delta m_{\rm spec} \sim -2.0$
mag.}
\label{syn:excess}
\end{figure*}

 What is the nature of the 132 sources in M\,83 that are known to host WR stars?
In Fig. \ref{syn:excess}(a) we compare the spectroscopic
  continuum magnitude to the spectroscopic excess, $\Delta m_{\rm spec} = $\he --
$m_{\lambda 4781}$, for all sources. This is more complete than  
Fig.\,\ref{photom:excess}, since it was generally possible to 
estimate the spectral \offhe \, magnitude for the fainter sources, where 
PSF-photometry was not available.

The brightest confirmed WR sources in our sample  (\offhe $\, \sim$20 mag)
exhibit --0.3 $\leq \Delta m_{\rm spec} \leq $0.0 mag.  Such
values are consistent with luminous complexes, greatly diluting the WR
emission signature. In contrast, the faintest confirmed sources
(\offhe $\, \sim$25 mag)  possess large spectroscopic excesses of --2 $\leq
\Delta m_{\rm spec} \leq$ --0.5 mag,  consistent with isolated,
single or binary WR systems. Intermediate
brightness sources span the full range in excess, corresponding
to less luminous regions hosting a few WR stars to those containing large
WR populations.

Fig. \ref{syn:excess}(b) compares the spectroscopic $\lambda 4686$
excess to the \Cthree$\lambda$4650/\He$\lambda$4686  equivalent width,
confirming the expected tight correlation between line strength and
$\Delta m_{\rm spec}$, where the scatter indicates the observational
accuracy. Typical excesses of --0.2 mag equate to small line
equivalent widths of $\sim$10\AA, whilst an excess of --1.0 mag
corresponds to $\sim$100\AA, and the largest excesses equate to
$\sim$500\AA. For comparison, single Galactic and LMC WR stars possess
 \blue\ equivalent widths of 10--500\AA\ (WN subtypes) or 150--2000\AA\
(WC subtypes).

\subsection{The global disk WR population of M\,83}

We identify 1035$\pm$300 WR stars, comprising 564$\pm$170 WC and 471$\pm$130 WN stars, within
our 132 spectroscopically observed regions, where errors quoted here
were obtained from simply adding individual uncertainties for all regions. 

The most important discovery of our spectroscopic survey is the
dominant late-type WC population of M\,83.  Over half of the
spectroscopically identified WR stars in M\,83 fall into the WC8--9
subtype, with few WC4--7 stars identified.  For comparison, no WC8--9
stars are observed in the SMC, LMC or M\,33 and the total number of
such stars in the Milky Way and M\,31 is less than 50
\citep{derhucht01, moffat87}.  The distribution among late- and
early-type WN stars is more even, with WNL/WNE $\sim 1$.  This value
is much greater than that observed in the SMC ($\sim$0) and LMC
($\sim$0.25), but comparable to that of $\sim 1.3$ determined for the
Milky Way \citep{derhucht01}.

How robust is this derived WR population for M\,83? For each source,
we have propagated uncertainties in the distance, reddening,
photometry and line flux measurements.  Together, these translate to a
typical uncertainty of $\sim 20 - 30$\%, or somewhat higher for
regions in which an interstellar reddening or a slit loss correction
factor have been adopted.

One of the main limitations in estimating the content of an unresolved
WR population is the conversion from WR line flux to WR content. Given
the large late WC population identified in M83, we have reconsidered
the line flux of individual WC8--9 stars determined by
\citet{schaerer}. From unpublished data for 5 Galactic, and 2
M\,31 WC8--9 stars, each with well derived distances, we find a mean
$\lambda 5696$ flux of $5.1 \times 10^{35} \mbox{ergs}^{-1}$ and $4.7
\times 10^{35} \mbox{ergs}^{-1}$ respectively. This is $\sim$30\%
lower than \citeauthor{schaerer}, and suggests that, if anything, we
may be underestimating the true WC population of M\,83.

We have also estimated the WC population using the alternative \Cthree
$\lambda 4650$ line.  Based on individual WR $\lambda 4650$ line
fluxes of $3.4 \times 10^{36},\, 4.5 \times 10^{36}\, \mbox{and}\, 1.0
\times 10^{35}\,\mbox{ergs}^{-1} \, $ for individual WC4--6, WC7 and WC8--9
subtypes, respectively \citep{schaerer},  populations of individual sources
were found to agree to within a factor of 2, relative to the yellow
features. The total WC population was calculated to be 594
using \Cthree $\lambda$4650, in excellent agreement with that of 564
obtained from \Cthree $\lambda 5696$ and C\,{\sc iv} $\lambda$5808.

Turning to the candidates for which spectroscopy was not obtained,
all regions in Fig.\,\ref{photom:excess} with $\Delta{\rm{m}} \leq$
--0.3 mag correspond to spectroscopically confirmed WR
  complexes.  Therefore,  we would expect that at least 25 out
of the 49  candidates, for which $\lambda$4684 and $\lambda$4781
photometry is available, also possess WR stars. Adopting the same
fraction for regions where PSF photometry is not available, we expect
$\geq$50 of the remaining 89 candidate regions to contain WR
stars.  Indeed, \#159 has already been observed by
\citet*{bresolin02}.  Designated M83-5 in their study, WR emission is
spectroscopically confirmed and a population of 2 WCL and 6 WNL stars
(scaled to a distance of 4.5Mpc) is inferred from its line
luminosity. On average, our confirmed sources host $\sim$5 WR stars,
such that we expect $\sim$250 WR stars await identification in M\,83,
bringing the total disk population to $\sim$1300.

The inferred WR population of M\,83 is greater than that known in
the entire Local Group, to date \citep{massey98}.  As anticipated from
Figs.\,\ref{photom:excess} and \ref{syn:excess}, some sources host
a single WR star, whilst  others contain
larger WR populations ($\sim 10$). Regions which contain an
exceptionally large WR population will be
discussed in more detail in the next section.


\subsection{Complexes hosting large WR populations}\label{clusters}

In the Milky Way, the most massive open clusters (e.g. Arches,
Westerlund 1) host at most 10--20 WR stars
\citep{blum01,negueruela05}.  Similar numbers are observed in the
largest H\,{\sc ii} regions of M\,33, and 30 Doradus in the LMC. We
identify 10 regions in M\,83 with large ($\geq$20), or mixed, WR
populations.  Mixed WN and WC populations are observed in a total of 5
complexes, \#66 (8$\pm$2 WNL, 4$\pm$1 WC7), \#38 (7$\pm$2 WNL,
21$\pm$6 WCL), \#41 (14$\pm$4 WNL, 13$\pm$6 WC7), \#86 (9$\pm$4 WNL,
24$\pm$10 WCL) and \#74 which will be discussed separately.

Are the sources that host WR stars in M\,83 compact clusters
(e.g. Arches) or extended, giant H\,{\sc ii} regions (30 Doradus)?
Massive compact clusters are generally rare in normal disk galaxies,
although M\,83 is known to host many examples, from HST imaging
\citep{larsen04}.  Of the 60 bright H\,{\sc{ii}} regions in M\,83
identified by \citeauthor{devau83},  between 28--38 host WR
populations.  Indeed, the 3 complexes hosting the largest WR
populations are all associated with H\,{\sc{ii}} regions identified by
\citeauthor{devau83}. Optical spectroscopy of these were presented in
Paper\,I, together with an estimate of their O star population.

\begin{figure}[ht!]
\includegraphics[width=7cm,angle=0]{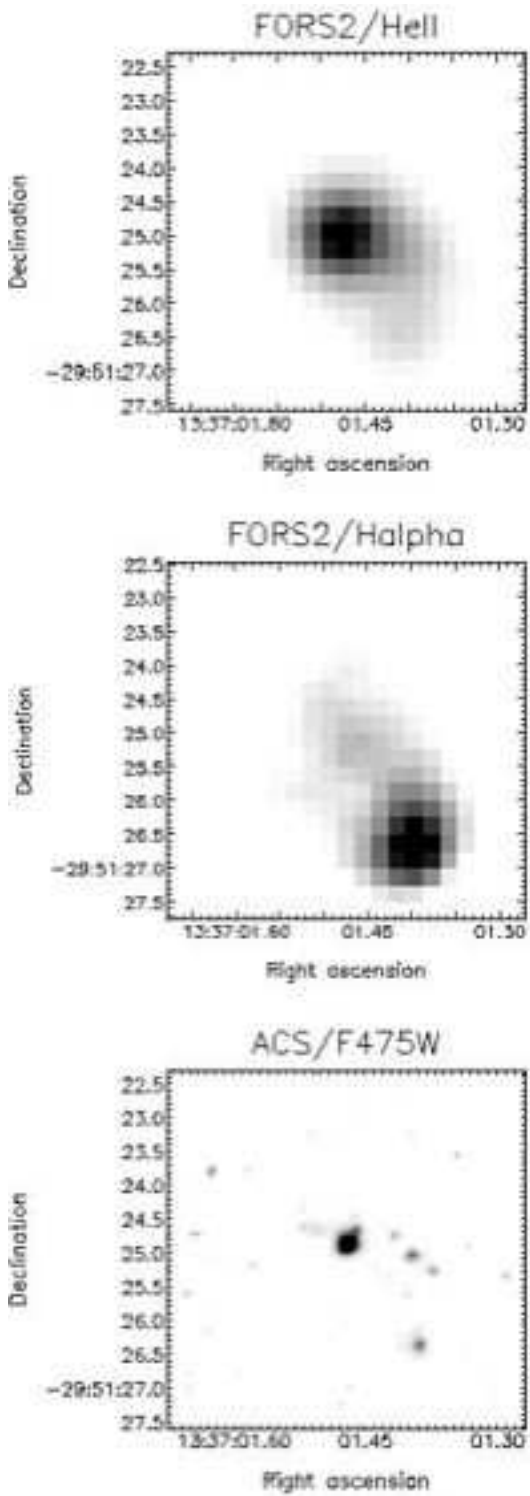}
 \caption{5$'' \times 5''$ images of the WR cluster M\,83 \#74 from VLT/FORS2 
and HST/ACS. Top Panel: $\lambda$4684 filter, middle panel: continuum 
subtracted H$\alpha$ filter, lower panel: F475W filter (WFC). North is up 
and east is to the left. It is apparent that the brightest H$\alpha$ source lies 
$\sim2''$ to the S-W from the continuum (and WR) source.}
 \label{74}
\end{figure}


\subsubsection{Source \#74}

From our sample \#74 is exceptional, with 230$\pm$50 late-type WN and
WC stars inferred from the de-reddened line fluxes (recall
Fig.\ref{fig:fits}).  This source has the highest interstellar
reddening of our entire sample with E(B-V)=1.0$\pm$0.03, although it
is closest to the nucleus.  However, the H$\alpha$/H$\beta$ nebular
value is supported from fitting its stellar continuum to a young
($\sim$4 Myr) instantaneous burst model at Z=0.04 from Starburst99
\citep{leitherer99}.  In Paper\,I, we estimated a Lyman continuum flux
of 8$\times 10^{51}$ s$^{-1}$ from the de-reddened H$\alpha$ flux,
such that \#74 has an ionizing flux equivalent to the giant H\,{\sc
ii} region 30 Doradus.  However, it possesses a WR content which is a
factor of ten times larger, i.e. N(WR)/N(O)$\sim$0.25 versus 0.02 in
30 Doradus.

We have inspected archival HST/Advanced Camera for Surveys (ACS) Wide
Field Camera (WFC) F475W datasets of M\,83 (Proposal 9299,
P.I. H. Ford).   This revealed that \#74 is very compact, with a FWHM of
$\sim$0.2 arcsec or $\sim$4.5\,pc (for a distance of 4.5Mpc).
 For H\,{\sc ii} regions with solar or super-solar
metallicities, WR signatures are expected to be present in bursts
of age 3--6Myr. We have compared the absolute F475W
magnitude of \#74 with evolutionary synthesis models for an
instantaneous burst of age 3--5Myr \citep{leitherer99}, from which we
estimate a mass of 1.4--2$\times 10^{5} M_{\odot}$.
Therefore, its mass {\it and} size indicate that it is a young massive compact 
cluster, or Super Star Cluster \citep{whitmore03}.  

In Fig.\ref{74} we present $5\times5$ arcsec  ($\sim110 \times 110$ pc) 
images of \#74 obtained
with FORS2 and ACS. It is apparent that the peak H$\alpha$ source,
i.e. H\,{\sc ii} region \#35 from \citet{devau83}, lies $\sim$2 arcsec
to the S-W of the brightest continuum source (the WR
cluster). The spectrum presented in Fig.2 of Paper\,I is that of the WR
cluster, whilst the H$\alpha$ flux, and corresponding O7V star content
of $\sim$810 represents the integrated total from both regions. The WR
cluster provides approximately 1/3 of the total H\,{\sc ii}
luminosity, such that the WR/O ratio of this region approaches unity,
comparable to the WR cluster NGC\,3125-1 \citep{chandar04}.

\subsubsection{Other Clusters in M\,83}

\citet{larsen04} has identified $\sim$80 young massive 
clusters in M\,83 based on HST/WFPC2 images.  Three 
such regions are in common with our catalogue of sources
containing WR stars, 
namely n5236-607 (\#61), -617 (\#73) and  -277 (\#79), 
although none host more than a few WR stars.
Larsen (priv. comm.) has compared the UBVI colours of these clusters
with Solar metallicity \citet{Bruzual93} models, suggesting age
estimates of $\log (\tau) = 6.20 \pm$0.51, 6.90$\pm$0.54 and 9.89$\pm$1.87,
respectively. The first two are fully consistent
with a young cluster which contains WR stars, while the third suggests
a dominant old population.

Five additional clusters from \citet{larsen04} are also in common with
our remaining candidates, namely n5236-169 (\#193), -805 (\#179), -818
(\#163), -1011 (\#157) and -1027 (\#173). Of course, such candidates
have the potential to also host a large WR population - indeed three
of these clusters appear young ($\sim$1.5--6Myr) from UBVI photometry
(Larsen, priv. comm.), i.e. \#193, \#179 and \#157. Note that 
\#179 is one of
two clusters for which dynamical mass estimates has been made by
\citet{larsen04b}. Follow-up spectroscopic observations would be
required for the identification of additional WR rich clusters.

\subsection{Comparisons with previous studies}

To date, there have only been two previous studies relating to
WR stars within M\,83. \cite{rosa87} and \cite{bresolin02} have both studied
stellar populations within M\,83 and identify six H {\sc ii} regions
which exhibit WR characteristics.  Four of these have been re-examined
in this study.  \citeauthor{rosa87} obtain optical spectra with very
poor signal-to-noise preventing a quantitative discussion,
consequently we shall restrict any comparisons solely to results
obtained by \cite{bresolin02}.  

Both studies followed a similar methodology in estimating the WR
population, except that \citeauthor{bresolin02} adopted a distance of
3.2Mpc to M\,83 (versus 4.5Mpc adopted here).  This introduces a
factor of 2 between intrinsic line luminosities observed in this study
and that by \citeauthor{bresolin02}.

\begin{itemize}
\item{\emph{ \#40 } (M83-2) --}
The derived WR population for this region is estimated to be 6$\pm$2
WC8--9, contrasting that of 1--2 WNL obtained by
\citeauthor{bresolin02}. We achieve a 3$\sigma$ detection for the
\redciii \, and \civ \, carbon features, suggesting that poor
signal-to-noise in the original investigation prevented positive WC
identification.
\item{\emph{ \#41} (M83-3) --}
We confirm the detection of 14$\pm$4 WNL stars identified in region
M83-3. In addition we estimate the presence of 13$\pm$6 WC7
stars. \citeauthor{bresolin02} state that \Cthree \, may be present,
but not at a significant level (versus $5 \sigma$ here).
\item{\emph{  \#74} (M83-8) --}
\citeauthor{bresolin02} failed to detect any WR emission in this
H\,{\sc ii} region.   However, we find the largest individual WR
population of any source, namely 230 stars.  As stated in
Sect.\ref{clusters} the WR emission is offset by several arc-secs to
the N-E of the peak \ha emission.  Since the \citeauthor{bresolin02}
concentrated on bright H {\sc{ii}} regions, their slit was probably
centred on the peak \ha emission, such that the WR signature was missed.
\item{\emph{  \#103} (M83-9) --}
Both investigations infer a late WN population.  The present
study obtains a population of 29$\pm$9 WNL stars, in agreement with that
estimated by \citeauthor{bresolin02} after allowing for
differences in the assumed distance.
\end{itemize} 

\begin{figure}[h!]
\centering
\includegraphics[width=7.8cm,angle=0]{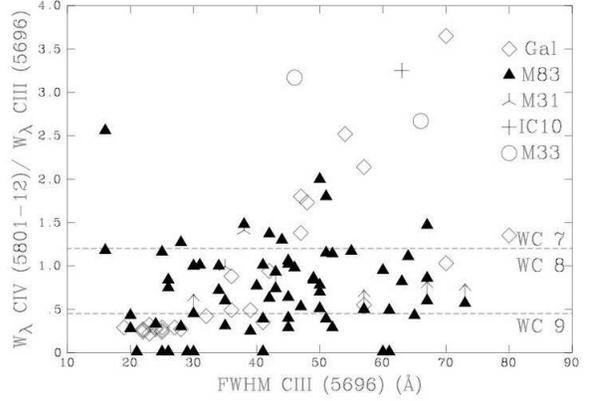}
 \caption{Distribution among WCL subtypes as determined using
 $\mbox{W}_{\lambda}$ (C\,{\sc iv} 5808) / $\mbox{W}_{\lambda}$
 (C\,{\sc iii} 5696) versus FWHM (C\,{\sc iii} $\lambda$5696) in
 \AA.  For comparison, Galactic (unpublished WHT, AAT and 2.3m ANU
 data), M31 (unpublished WHT / ISIS data), M33 \citep{abbott04} and
 IC10 \citep{pac03} WCL stars have been included. The subtype
 divisions marked are those derived by \citet{pac98}.}
\label{WCL}
\end{figure}

\begin{figure}[h!]
\centering
\includegraphics[width=7.8cm,angle=0]{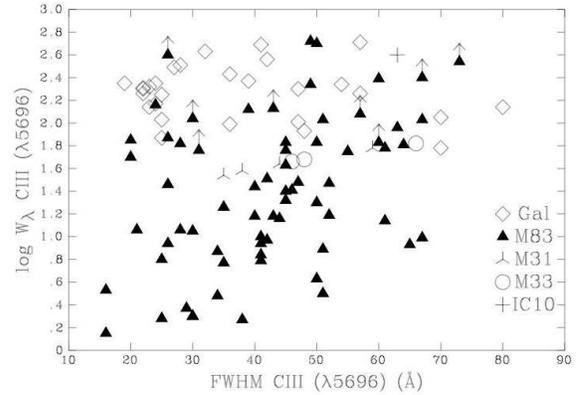}
 \caption{Equivalent width ($\mbox{W}_\lambda$ in \AA) vs. FWHM (\AA) \,of the \Cthree  
$\lambda 5696$ for WCL stars in
M\,83 and Local Group galaxies (identical dataset as presented in Fig.\ref{WCL}).}
\label{WCL:EW}
\end{figure}

\section{Discussion}\label{discussion}

 We have identified up to $\sim$200 regions in the disk of M\,83
that host WR stars. We now compare the properties of
WR  stars at the high metallicity of M\,83 with 
those of Local Group galaxies, attempt to
explain the dominant late subtypes amongst WC stars, and make comparisons
with current evolutionary models.

\subsection{Properties of WR stars at high metallicity}

How do the line strengths and widths of sources containing WR stars in M\,83 compare 
with those of other galaxies? In Fig.\,\ref{WCL},  we show the classification ratio
$\mbox{W}_\lambda$ (C\,{\sc iv} 5808) / $\mbox{W}_\lambda$ (C\,{\sc iii}
5696) versus FWHM (\Cthree\, 5696). Data for WC7--9 stars in four Local Group 
galaxies are included, along with 
subtype boundaries as derived by \citet{pac98}.  This figure highlights
the dominance of WC8 and WC9 subtypes,
which comprise 95\% of the total WC content of M\,83, by number.

The Galactic WC9  population is very homogeneous,
centred on a FWHM of $\sim$25\AA\, and $\mbox{W}_\lambda$ (C\,{\sc iv}
5808) / $\mbox{W}_\lambda$ (C\,{\sc iii} 5696) $\sim$0.3.  In contrast,
the WC9 population of M\,83 is very heterogeneous, spanning a wide
range of both FWHM and $\mbox{W}_\lambda$ (C\,{\sc iv} 5808) /
$\mbox{W}_\lambda$ (C\,{\sc iii} 5696), with the maximum FWHM reaching
$\sim 60$\AA, 2.5 times greater than the typical Galactic WC9
star. Line widths of  sources hosting WC8 stars  are much greater
than typical Galactic WC8 stars, whilst the few sources containing
WC7 subtypes are more indicative of Milky Way counterparts.

In Fig.\,\ref{WCL:EW} we present \Cthree $\lambda$5696 line width
versus line strength for WCL stars observed in M\,83, along with data
for single or binary WCL stars in Local Group galaxies. The majority
of WR complexes observed in M\,83 display evidence for line dilution
from underlying stellar continua, since line strengths fall well below
those observed in Local Group counterparts. Some WR complexes in M\,83
do display similar line strengths to those in the Milky Way or
M31/M33, suggesting little evidence of line dilution in those cases.

\begin{figure}[h!]
\centering
\includegraphics[width=7.8cm,angle=0]{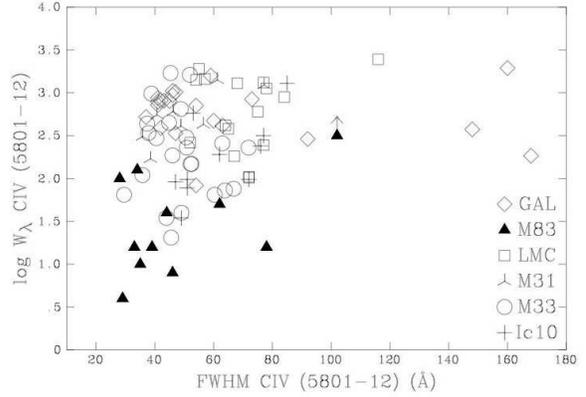}
 \caption{Equivalent width ($\mbox{W}_\lambda$ in \AA) and FWHM (\AA) of the \Cfour
 $\lambda 5808$ line for early WC stars. Galactic, LMC, M\,31, M\,33 and
 IC10 WCE (unpublished WHT, AAT and 2.3m ANU data) and WO
 \citep{drew04,kingsburgh95} stars are also indicated.}
 \label{WCE}
\end{figure}

In Fig.\,\ref{WCE} we compare $\mbox{W}_{\lambda}$ (\Cfour $\lambda
5808$) to FWHM (\Cfour $\lambda 5808$) for early-type WC stars in
M\,83, with (mostly single) Galactic and LMC counterparts.  Again,
 the observed line strengths of sources containing early WC stars
in M\,83  fall below those of Galactic and LMC stars.   This is again
attributed to line dilution by the underlying continua from early-type 
stars.

From Fig.\,\ref{WCE}, the observed WCE line widths of M\, 83 members are
generally comparable to, or lower than, those of other Local Group
members, in contrast with WC8, and especially, WC9 subtypes. Indeed,
there are no cases for which FWHM (C\,{\sc iv} $\lambda$5808) $\geq$
100\AA, corresponding to WO subtypes in the Milky Way/LMC
\citep{kingsburgh95, drew04}, except possibly \#6  for which no evidence
of O\,{\sc vi} $\lambda$3811-34 is observed.
%

\begin{figure}[htbp]
 \centering
\includegraphics[width=7.8cm,angle=0]{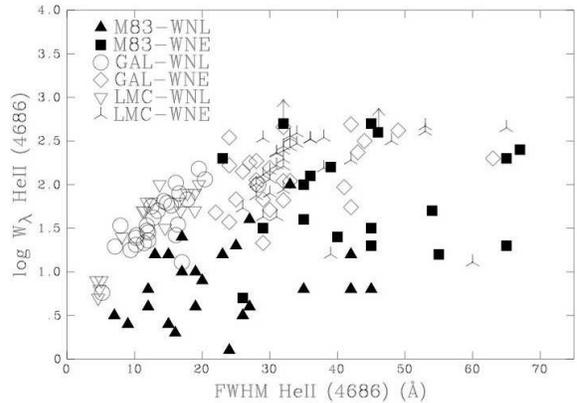}
 \caption{A comparison between the equivalent width
 ($\mbox{W}_\lambda$ in \AA) of the \He $\lambda 4686$ line and its FWHM (\AA). WN
 stars in M\,83 and Local Group galaxies are marked, populations have been
 divided into WNE and WNL subtypes.  Data for Galactic and LMC WNL stars
 are taken from \citet{pac97}, WNE information can be found in \citet{conti89}.}
 \label{WN}
\end{figure}

Finally, in Fig.\ref{WN} we compare the equivalent width and FWHM of
\He $\lambda$4686 for all WN sources identified in M\,83.  Again, we have
included data for single/binary Galactic and LMC WN stars.  Aside from the effect of
line dilution, some late-type WN stars in M\,83 possess unusually large
line widths.   In some M83 complexes hosting multiple WN stars, line
widths are a factor of two greater than Galaxy or LMC counterparts.
These are reminiscent of  observations of broad, strong \Nthree $\lambda$4640
in unresolved WR galaxies \citep{schmutz99}.


\subsection{Origin of late WC stars at high metallicity?}

\begin{figure}[htbp]
 \centering
\includegraphics[width=6.5cm,angle=0]{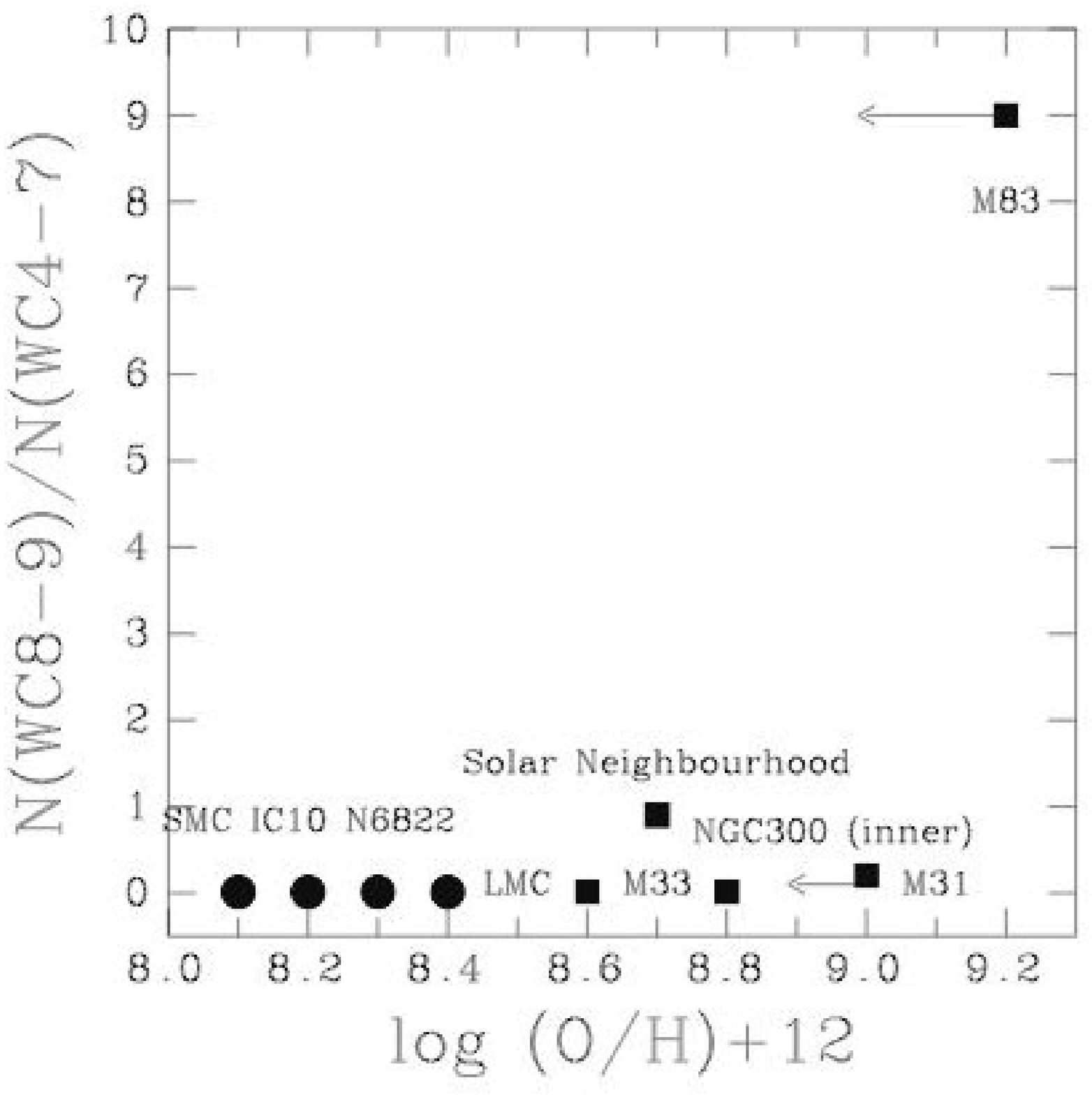}
 \caption{The fractional distribution of late to early WC stars in
M\,83 and Local Group galaxies versus metallicity from
\citet{massey98}, \citet{schild03}, \citet{pac03} and the present
study. The oxygen content of M\,83 is taken from \citet{bresolin02},
although more recent results suggest a reduction by $\sim$0.2 dex
\citep{pilyugin04, bresolin04}}
 \label{WCL:WCE}
\end{figure}

In Fig.\,\ref{WCL:WCE} we compare the fractional distribution of WC8--9 to WC4--7
stars in galaxies with a wide range of metallicity. This clearly
illustrates the extreme WCL population hosted by M\,83, indicating that
WCL stars are uniquely associated with metal-rich environments.  In
M\,83 the relative number of late to early WC stars is found to be
$\sim 9$, much greater than 0.9 and $\sim 0.2$ observed for the Milky
Way and M\,31 respectively.  C\,{\sc iii} $\lambda$5696 has been
observed in a small number of metal-rich WR galaxies
\citep{phillips92, pindao02}, but as these represent integrated
populations, the
`average' WC subtype is generally WC7--8.

It has long been recognised that Milky Way WC9 stars are
universally observed towards the Galactic Centre.
\citet{smith91} argued that the apparent trend towards later subtypes
was due to heavy mass-loss, revealing WC subtypes at an earlier
evolutionary phase, {\it assuming} the surface (C+O)/He ratio
decreases 
from early to
late WC subtypes. However, subsequent spectral analysis failed to
confirm any systematic trend in C/He with subtype \citep{koesterke95},
arguing against late WC stars being  exposed earlier due to prior
mass-loss.

Instead, \citet{pac02} claimed that WC subtypes resulted from
primarily metallicity-dependent wind strengths.  They suggested that
the strength of C\,{\sc iii} $\lambda$5696 scales very sensitively
with wind density. If wind strengths increase with increasing (heavy
element) metallicity, as already established for OB stars, stars which
are otherwise identical will only reveal strong C\,{\sc iii}
$\lambda$5696 emission at high metallicities, with a corresponding
late subtype.

 Indeed, late-type WC stars are observed across the disk of M\,83 at a 
lower average galactocentric distance of 2.9 $\pm$ 0.9 arcmin ($\rho$ = 
0.4 $\pm$ 0.1 $\rho_{0}$) versus 3.6 $\pm$ 0.9 arcmin ($\rho$ = 0.5 $\pm$
0.1 $\rho_{0}$) for early-type WC stars. This can be explained by the
weak metallicity gradient observed in M\,83 \citep{pilyugin04} since stars
at smaller galactocentric distances will be more metal-rich
than those at larger galactocentric distances.

As discussed in the introduction, the  lower mass limit for stars
that ultimately become WR stars decreases with increasing metal content, i.e.
the lifetimes of low (initial) mass WR stars are greatly enhanced
relative to those at lower metallicity. Could late WC stars be the descendants
of such  (low initial mass) stars, such that they greatly outnumber the (initially
more  massive) early WC stars? 

 In this scenario, one would expect late WC stars to be observed in
Milky Way clusters with low mass turn-off's.  \citet{schild86}
identified the Galactic WC stars WR77 (WC8) and WR95 (WC9) in open
clusters with progenitor masses as low as 35$M_{\odot}$, whilst early
WC stars appear to originate from more massive progenitors ($\geq$
60$M_{\odot}$).  More recently, \citet{massey01} identified the late
WC WR93 (WC7) with a very high progenitor mass of $\geq 120
M_{\odot}$, comparable to or higher than early WC subtypes.
Unfortunately, WC8--9 stars were not included in their study.  In
addition, the late WC component of WR11 ($\gamma$ Vel, WC8+O7.5)
originates from a mass somewhat in excess of 30$M_{\odot}$, the
current mass of the O star companion \citep{demarco99}.  Overall,
there is limited evidence for a distinction between the progenitor
masses of early and late WC stars.  Since nebular H$\alpha$ emission
scales inversely with age, the complexes hosting early- and late-type
WC stars appear to be located in both young and old clusters, such
that they do indeed originate from the same parent population.

Consequently, the observed WC population in M\,83 can most naturally be
explained if WC winds are metallicity dependent. Increased
mass-loss would elevate the strength of the classification line
C\,{\sc iii} $\lambda$5696 resulting in predominantly late WC
subtypes.  Increased mass-loss rates would, of course, have
implications for the life-times of WC stars. For an initial mass of 40
M$_{\odot}$ and $Z=0.04$, a WR star would spend 50\% less time in the
WC phase with a metal dependent stellar wind \citep[See][]{meynet04}.
At the current stage, however, it is not possible to firmly exclude different
progenitor masses for early and late WC populations.

Finally, a similar metallicity effect was earlier proposed by
\citet{pac98b}, i.e.  WO subtypes would be restricted to environments
with weak winds (i.e. low metallicities) due to the inverse
sensitivity of O\,{\sc vi} $\lambda$3811-34 with increased
mass-loss. The absence of WO stars in M\,83 is also consistent with
the inverse sensitivity of the classification line O\,{\sc vi}
$\lambda$3811-34 to increasing wind strength due to higher
metallicity.

\begin{table}[t!]
\caption{The WR population of the LMC, Milky Way and M\,83. The 
cluster/association hosting the largest
WR population is also indicated.}
\label{large:clus}
\begin{tabular}{l@{\hspace{-3mm}}c
@{\hspace{1.5mm}}c@{\hspace{1.5mm}}c@{\hspace{1.5mm}}c@{\hspace{1.5mm}}r}
\hline
Galaxy &  log(O/H) & N(WN)& N(WC)&N(WR)&WC/WN\\
\hline
LMC$\,^{(1)}$ &8.4& 109& 24&134&0.2\\
-- 30Dor$\,^{(1)}$&&15&3&18&\\
Milky Way$\,^{(2)}$ & 8.7 & 132&92&237&0.7\\
-- Arches$\,^{(3)}$&&15&0&15&\\
-- Wd1$\,^{(4)}$& &$\geq$ 12& $\geq$ 7& $\geq$ 19&\\
M\,83&9.2&471$\pm$130&564$\pm$170&1035$\pm$300&1.2\\
-- \#74 &&52$\pm$12&179$\pm$42&231$\pm$50&\\
\hline
\hline
\end{tabular}
$\,^{(1)}$\,\citet{breysacher99},
 $\,^{(2)}$ \citet{derhucht01}, $\,^{(3)}$ \citet{blum01}, $\,^{(4)}$
\citet{negueruela05}.
\end{table}

\subsection{Comparison with evolutionary predictions}

Surveys for WR stars in Local Group galaxies over the past three
decades have revealed a strong correlation between the relative number
of WC to WN stars and oxygen content of the host galaxy
\citep{massey96}.  Extrapolating from previous observations, one would
expect N(WC)/N(WN) $\geq 1$ for a galaxy forming stars continuously
with $\sim$twice the Solar oxygen content \citep{massey98}.

\begin{figure}[h!]
 \centering
\includegraphics[width=6cm,angle=0]{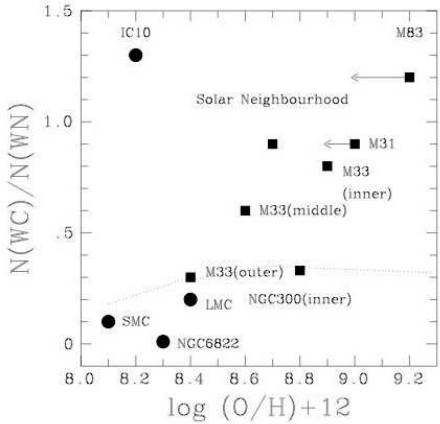}
 \caption{The WC/WN Ratio: Relative number of WC to WN stars in M\,83 is
 compared with those for nearby galaxies as a function of oxygen abundance,
as  determined by \citet{massey98}, \citet{schild03}, \citet{pac03}. We include
recent evolutionary predictions for rotating massive stars
from \citet{meynet04} (dotted line).} 
 \label{WC:WN}
\end{figure}

A summary of results for the disk population of  M\,83 are presented in 
Table \ref{large:clus} together with
Local Group members, such that the observed N(WC)/N(WN) ratio is presented 
in Fig.\ref{WC:WN}. Indeed, M\,83 continues
the observed trend rather well with N(WC)/N(WN)$\sim$1.2. Undoubtedly, 
completeness should obviously be kept in mind given that WC stars are
more readily identified in external galaxies than WN stars due to their 
intrinsically stronger lines. Nevertheless, our approach is optimised for 
net emission at $\lambda$4686, such that we achieve cases of
4$\sigma$ spectroscopic WNL detections with $W_{\lambda}$(He\,{\sc 
ii} $\lambda$4686) $\sim$ 1\AA. 


Recent bursts of star formation may cause strong deviations
from the general trend via a strong enhancement of the WC population at
an age of $\sim$5Myr \citep{pindao02}. Consequently, galaxies in which
there is a significant recent starburst episode may strongly deviate
from this correlation.  IC\,10 strongly deviates from
 the overall trend in Fig.\,\ref{WC:WN} due to an apparent 
`galaxy-wide' starburst \citep{pac03}, although the WN population 
may be significantly incomplete \citep{massey02}. 

In the case of M\,83, since we have no information on the nuclear
starburst, the above statistics should be reasonable for the quiescent
star forming regions, except that the presence of a recent starburst
in \#74 represents a non-negligible fraction of the total disk WR
population. If we were to exclude \#74 from our statistics, we would
obtain N(WC)/N(WN)=1 (equivalent to the Milky Way) and
N(WC8--9)/N(WC4--7)=6 (vs 0.9 in the Milky Way) for the quiescent
disk. The subtype ratios remain far from current evolutionary
predictions at high metallicity \citep{meynet04}.

Recently, \citet{meynet04} have constructed a set of evolutionary
models for rotating (initially $v_{\sin i}=300$ km\,s$^{-1}$) massive
stars from $Z$=0.004 (SMC) to 0.04 ($\sim$M\,83). For low metallicity
Local Group galaxies, predictions from rotating models are in good
agreement with observed WC to WN ratios (see
Fig.\,\ref{WC:WN}). However, at higher metallicities, even allowing
for metallicity dependent WR winds, evolutionary models fail and
dramatically underestimate the number of WC stars, i.e. WC/WN=0.36 at
$Z$=0.04.  In fact, non-rotating models provide a better match to our
observations of M\,83, although such models are unsuccessful in
predicting the correct fraction of Type Ib/c to Type II Supernovae at
high metallicity. Consequently, there remains a significant
discrepancy between the observed and predicted WR populations above
Solar metallicities.  

Evolutionary models distinguish between late and early WN subtypes via
the presence or absence of hydrogen, whilst spectroscopic definitions
relate to the observed ionization of nitrogen lines. Determinations of
hydrogen content are possible for Local Group WN stars, but the strong
nebulosity and potential multiplicity for sources at the distance of
M\,83 prevent such measurements.  For a twice Solar metallicity
($Z$=0.04), \citet{meynet04} predicts WNL/WNE $\sim$ 4 by number,
allowing for metallicity dependent WR wind strengths. This should
provide a reasonable analogue to the observational statistics for
M\,83, on the basis of a good correspondence between late WN stars
(with hydrogen) and early WN stars (without hydrogen) in the Milky
Way. Consequently, as with the WC to WN ratio, the predicted
distribution amongst WN subtypes ($\sim$4) differs from observations
($\sim$1) by a significant factor.

Nitrogen in WN stars is of course partially processed from carbon and
oxygen, such that the CNO equilibrium abundance linearly scales
with metallicity.  As a consequence, the abundance of nitrogen in M\,83
WN stars will exceed that in the Milky Way and other Local Group
galaxies. \citet{pac00} demonstrated that for otherwise identical
parameters, N\,{\sc iii} $\lambda$4634--41 reacts more sensitively
than N\,{\sc iv} $\lambda$4058 to increased nitrogen content, i.e. a
later WN subtype results. If WN winds are also metallicity dependent,
the effect will be magnified such that one will expect a
predominantly late WN population at high metallicities.

Indeed, given the weak metallicity gradient of M\,83  \citep{pilyugin04},
late-type WN stars are located within the inner regions of
M\,83, at an average galactocentric distance of 2.2$\pm$0.9 arcmin
($\rho$ = 0.3$\pm$0.1 $\rho_{0}$) versus 3.5$\pm$1.3 arcmin
($\rho$ = 0.5$\pm$0.2 $\rho_{0}$) for early-type WN stars. 

\section{Summary}\label{conc}

Our analysis of the VLT/FORS2 imaging and spectroscopic data indicates
that the disk of M\,83 hosts a large WR population.  Using narrow-band
optical images we have identified 283 candidate WR regions within
M\,83, of which 198 have been spectroscopically observed.  Of these we
find that 132 regions contain WR stars.  Absolute WR populations have
been derived using line flux conversions adapted from
\citet*{schaerer}.  We estimate a total WR population of 1035$\pm$300,
consisting of 564$\pm$170 WC and 471$\pm$130 WN stars from this population, i.e.  a
quiescent N(WC)/N(WN) ratio of $\sim$1.2, or $\sim$1.0 excluding the
starburst cluster \#74. This differs greatly from current evolutionary
predictions at high metallicity, which suggest N(WC)/N(WN)$\sim$0.36,
even allowing for metallicity dependent WR mass-loss rates
\citep{meynet04}.

The observed statistics exclude both the potentially large WR
population in the central starburst, plus the population of
perhaps $\sim$250 WR stars resulting
from remaining candidates. \citet{pellerin04} has carried out spectral
synthesis of $FUSE$ observations of the nucleus of M\,83(30$''\times 30''$)
suggesting a mass of 1.5$\times 10^{6} M_{\odot}$ and age 3.5 Myr,
with an inferred WR population of 1700. This appears
plausible given that the nuclear starburst has a star formation rate
approaching that of the disk \citep{harris01, bell01}. Consequently,
the total WR population of M\,83 may exceed 3000. 

Using the WR population derived in this study, the global surface
density of M\,83 is found to be $\sim$3 WR/kpc$^{2}$, typical of that
observed in the Solar Neighbourhood and M33 \citep{massey98}.  If the
nucleus and remaining candidates are accounted for this likely to
increase this by a factor of $\sim$3.


The WC population of M\,83 is dominated by late-type stars. The
relative number of WC8--9 to WC4--7 stars is found to be $\sim 9$ (or
$\sim$6 excluding cluster \#74), outnumbering that of any other Local
Group galaxy tenfold, as illustrated in Fig. \ref{WCL:WCE}.  WO
stars are not observed in M\,83 suggesting that there is a genuine trend
to later subtype at higher metallicities. Observed line widths in
early WC subtypes appear to be comparable to those observed in the
Milky Way and LMC. This population is most readily explained by a
metallicity dependent wind strength amongst WC subtypes.

M\,83 has a substantial WN population, evenly split between early
and late subtypes. The high WNL population likely results from
the sensitivity of nitrogen diagnostics to the high global metallicity, 
whilst evidence in favour of metallicity dependent winds amongst WN
subtypes is less clear. Evolutionary  models, in contrast, predict
a far higher late WN population at high metallicities, with WNL/WNE$\sim$4--5
\citep{meynet04}.

From the present study we infer 9  complexes in M\,83 which
contain a large WR population ($> 20$), with \#74 hosting over 200
late-type WN and WC stars, outnumbering 30 Doradus tenfold.  HST/ACS
images indicate that \#74 is a compact cluster, whilst several other
massive compact clusters from \citet{larsen04} host more modest WR
populations.

Within galaxies located at up to $\sim$10\,Mpc, how unusual are the
sources in M\,83 with regard to WR content?  \citet{schaerer99}
identified $\sim$40 WN and WC stars in clusters A and B of NGC\,5253,
similar to \#41 in M\,83, whilst He\,2--10 dwarfs even \#74, with 1100
WN and $>$250 WC stars.  Recently, \citet{chandar04} use HST/STIS UV
spectroscopy to claim NGC\,3125-1 hosts 5000 WNL stars, a factor of ten
times greater than optically derived by \citet{schaerer99}, and
comparable to the global WR population in M\,83.  Consequently, the WR
population of M\,83, and source \#74 in particular, is extreme only
with respect to Local Group galaxies. Where our results for M\,83 stand
out from previous studies is the ability to resolve the disk WR
population into $\sim$200 regions.

Given such a large WR population in the
optically visible disk of M\,83, it is likely that one such massive star 
will undergo a core-collapse each century, given typical WR lifetimes of
$\sim$10$^{5}$ yr. Indeed, SN1983N \citep{porter87}, one of
the six optically visible SN that have been reported in M83 since
1923, was a type Ib SN, for which WR stars are considered as
the most probable precursors.

In order to directly witness a WR explode as a supernova on a shorter
timeframe of a decade or so, a survey of the WR population in 10--20
nearby ($<$10Mpc) massive star forming galaxies would be required.
The present observational program will continue towards this goal,
complementing existing broad-band pre-Supernova surveys
\citep{smartt03, vandyk03}.

\begin{acknowledgements}
LH acknowledges financial support from PPARC and a Royal Society
summer studentship award, PAC acknowledges financial support from the
Royal Society.  We thank Soeren Larsen for providing ages estimates of
clusters from broad-band photometry.  Some of the data presented in
this paper were obtained from the Multimission Archive at the Space
Telescope Science Institute (MAST). STScI is operated by the
Association of Universities for Research in Astronomy, Inc., under
NASA contract NAS5-26555. Support for MAST for non-HST data is
provided by the NASA Office of Space Science via grant NAG5-7584 and
by other grants and contracts.
\end{acknowledgements}

\appendix

\section{Catalogue}

Tables A1 and A2 list properties of confirmed and candidate regions containing
WR stars in M\,83.

\section{Finding charts (electronic only)}

Fig. B1 provides a master key to individual finding charts B2--B17, 
obtained from FORS2 $\lambda$4684 images.


\begin{sidewaystable*}[h]
{\bf{Table A1.}} Catalogue of sources with confirmed WR spectroscopic
signatures in M\,83.  Errors are indicated on the second row for each
source. Note $\Delta m$ and M$_{\rm B}$ shown in parentheses
correspond to spectroscopic photometry.  De-projected distances are
expressed as a fraction of the Holmberg radius $\rho_{0} = 7.3 ' =
9.56\,\mbox{kpc}$, equivalent widths $\mbox{W}_\lambda$ and FWHM are
in \AA\,and observed line fluxes, $F_{\lambda}$, are expressed in
$\mbox{erg s}^{-1} \mbox{cm}^{-2}$.  The number of, and error in, WR
stars in each source are estimated from the line luminosities
indicated in Sect 3.2 and a distance of 4.5$\pm$0.3\,Mpc
\citep{thim03}. \\
\begin{tabular}{c
@{\hspace{1mm}}c
@{\hspace{2mm}}c
@{\hspace{2mm}}c
@{\hspace{2mm}}c
@{\hspace{1mm}}c
@{\hspace{1mm}}l
@{\hspace{1mm}}c
@{\hspace{1mm}}c
@{\hspace{1mm}}r 
@{\hspace{2mm}}c 
@{\hspace{0.5mm}}c 
@{\hspace{1mm}}r 
@{\hspace{2mm}}c 
@{\hspace{0.5mm}}c 
@{\hspace{1mm}}r 
@{\hspace{2mm}}c 
@{\hspace{0.5mm}}c 
@{\hspace{0mm}}c 
@{\hspace{1.5mm}}c
@{\hspace{2.5mm}}c
@{\hspace{1.5mm}}c
@{\hspace{1.5mm}}c
@{\hspace{1.5mm}}c 
@{\hspace{0.5mm}}
}
\noalign{\hrule\smallskip}
Cat & RA & Dec& $\rho/\rho_0$ & $m_{4686}$ & $\Delta m$ & $E_{\rm B-V}$ & $M_{\rm B}$ & DeV & \multicolumn{3}{c}{C\,{\sc iii} $\lambda$4650/He\,{\sc ii} 4686} & 
\multicolumn{3}{c}{C\,{\sc iii} $\lambda$5696} & 
\multicolumn{3}{c}{
C\,{\sc iv} $\lambda$5808} & \multicolumn{2}{c}{WN} 
& \multicolumn{3}{c}{WC} & MOS\\   
\# & \multicolumn{2}{c}{J2000} &  & mag & mag & mag & mag &
& $W_{\lambda}$ & FWHM & $F_{\lambda}$ 
& $W_{\lambda}$ & FWHM & $F_{\lambda}$ 
& $W_{\lambda}$ & FWHM & $F_{\lambda}$ & 2--5 & 6--10 & 4--6 & 7 &
8--9 & Mask \\
\noalign{\hrule\smallskip}
1\factor&13:36:40.28&-29:51:21.1&0.65&\multicolumn{2}{c}{Extended}&0.50&&1&3&15& 3.8$\times 10^{-16}$ && &&& & & &3& & & &D5,B5\\
&&&& &  &  0.01& & &&& 1.5$\times 10^{-16}$&&& &&& & & 1& & & &\\
2&13:36:41.04&-29:51:55.9&0.63& 18.2&  -0.2&  0.54& -11.8& &6&15& 6.1$\times 10^{-16}$&1&15& 2.5$\times 10^{-16}$&2&35& 2.9$\times 10^{-16}$& & & & 2& &D4\\
&&&& 0.0&  0.1&  0.01& 0.2& &&& 2.9$\times 10^{-16}$&&& 7.0$\times 10^{-17}$&&& 9.1$\times 10^{-17}$& & & & 1& &\\
3\factor&13:36:41.14&-29:51:48.8&0.62&(24.6)&(-1.9)& 0.41&(-5.3)&&$>$400&20& 1.8$\times 10^{-16}$&$>$400&25& 2.4$\times 10^{-16}$&$>$300&20& 5.6$\times 10^{-17}$& & & & & 2&D2\\
\end{tabular}
\begin{flushleft}
\begin{small}
\factor)~\He PSF or spectroscopic photometry was unavailable and the average slit loss correction factor has been assumed,
\red)~the average E(B--V) value has been assumed,
\extended)~photometry corresponds to a lower limit as the region is extended.  
\end{small}
\end{flushleft}
\end{sidewaystable*}
\begin{sidewaystable*}[h]
{\bf{Table A1.}} (continued)\\
\begin{tabular}{c
@{\hspace{1mm}}c
@{\hspace{2mm}}c
@{\hspace{2mm}}c
@{\hspace{2mm}}c
@{\hspace{1mm}}c
@{\hspace{1mm}}l
@{\hspace{1mm}}c
@{\hspace{1mm}}c
@{\hspace{1mm}}r 
@{\hspace{2mm}}c 
@{\hspace{0.5mm}}c 
@{\hspace{1mm}}r 
@{\hspace{2mm}}c 
@{\hspace{0.5mm}}c 
@{\hspace{1mm}}r 
@{\hspace{2mm}}c 
@{\hspace{0.5mm}}c 
@{\hspace{0mm}}c 
@{\hspace{1.5mm}}c
@{\hspace{2.5mm}}c
@{\hspace{1.5mm}}c
@{\hspace{1.5mm}}c
@{\hspace{1.5mm}}c
@{\hspace{0.5mm}}
}
\noalign{\hrule\smallskip}
Cat & RA & Dec                 & $\rho/\rho_0$ & $m_{4686}$ & $\Delta m$ 
& $E_{\rm B-V}$ & $M_{\rm B}$ & DeV & \multicolumn{3}{c}{C\,{\sc iii} $\lambda$4650/He\,{\sc ii} 4686} & 
\multicolumn{3}{c}{C\,{\sc iii} $\lambda$5696} & 
\multicolumn{3}{c}{
C\,{\sc iv} $\lambda$5808} & \multicolumn{2}{c}{WN} 
& \multicolumn{3}{c}{WC} & MOS\\   
\# & \multicolumn{2}{c}{J2000} &  & mag & mag & mag & mag &
& $W_{\lambda}$ & FWHM & $F_{\lambda}$ 
& $W_{\lambda}$ & FWHM & $F_{\lambda}$ 
& $W_{\lambda}$ & FWHM & $F_{\lambda}$ & 2--5 & 6--10 & 4--6 & 7 &
8--9 & Mask \\
\noalign{\hrule\smallskip}
21& 13:36:48.54& -29:51:11.4& 0.39&  21.9&   -0.6&   0.20&  -6.7&  & 70& 30&  4.7$\times 10^{-16}$& 45& 45&  2.0$\times 10^{-16}$& 45& 25&  1.6$\times 10^{-16}$&  &  &  &  &  1& D2\\
& & & &  0.0&   0.1&   0.05&  0.3&  & & &  8.7$\times 10^{-17}$& & &  1.0$\times 10^{-16}$& & &  4.1$\times 10^{-17}$&  &  &  &  &  1& \\
22& 13:36:48.90& -29:51:44.2& 0.37&  22.2&  (-1.5)&  0.19&  -5.7&  & $>$350& 45&  4.8$\times 10^{-16}$& & &  & & &  &  4&  &  &  &  & B1\\
& & & &  0.1&   &   0.04&  0.2&  & & &  4.7$\times 10^{-17}$& & &  & & &  &  1&  &  &  &  & \\
\end{tabular}
\begin{flushleft}
\begin{small}
\factor)~\He PSF or spectrscopic photometry was unavailable and the average slit loss correction factor has been assumed,
\red)~the average E(B--V) value has been assumed,
\extended)~photometry corresponds to a lower limit as the region is extended.  
\end{small}
\end{flushleft}
\end{sidewaystable*}
\begin{sidewaystable*}[h]
{\bf{Table A1.}} (continued)\\
\begin{tabular}{c
@{\hspace{1mm}}c
@{\hspace{2mm}}c
@{\hspace{2mm}}c
@{\hspace{2mm}}c
@{\hspace{1mm}}c
@{\hspace{1mm}}l
@{\hspace{1mm}}c
@{\hspace{1mm}}c
@{\hspace{1mm}}r 
@{\hspace{2mm}}c 
@{\hspace{0.5mm}}c 
@{\hspace{1mm}}r 
@{\hspace{2mm}}c 
@{\hspace{0.5mm}}c 
@{\hspace{1mm}}r 
@{\hspace{2mm}}c 
@{\hspace{0.5mm}}c 
@{\hspace{0mm}}c 
@{\hspace{1.5mm}}c
@{\hspace{2.5mm}}c
@{\hspace{1.5mm}}c
@{\hspace{1.5mm}}c
@{\hspace{1.5mm}}c
@{\hspace{0.5mm}}
}
\noalign{\hrule\smallskip}
Cat & RA & Dec                 & $\rho/\rho_0$ & $m_{4686}$ & $\Delta m$ 
& $E_{\rm B-V}$ & $M_{\rm B}$ & DeV & \multicolumn{3}{c}{C\,{\sc iii} $\lambda$4650/He\,{\sc ii} 4686} & 
\multicolumn{3}{c}{C\,{\sc iii} $\lambda$5696} & 
\multicolumn{3}{c}{
C\,{\sc iv} $\lambda$5808} & \multicolumn{2}{c}{WN} 
& \multicolumn{3}{c}{WC} & MOS\\   
\# & \multicolumn{2}{c}{J2000} &  & mag & mag & mag & mag &
& $W_{\lambda}$ & FWHM & $F_{\lambda}$ 
& $W_{\lambda}$ & FWHM & $F_{\lambda}$ 
& $W_{\lambda}$ & FWHM & $F_{\lambda}$ & 2--5 & 6--10 & 4--6 & 7 &
8--9 & Mask \\
\noalign{\hrule\smallskip}
41\factor& 13:36:52.64& -29:51:47.5& 0.26& \multicolumn{2}{c}{Extended}&  0.52& & 12& 5& 20&  1.5$\times 10^{-15}$& 2& 30&  1.4$\times 10^{-15}$& 2& 45&  1.9$\times 10^{-15}$&  &  14&  &  13&  & B3\\
& & & &  &   &   0.02&  &  & & &  3.9$\times 10^{-16}$& & &  3.5$\times 10^{-16}$& & &  9.0$\times 10^{-16}$&  &  4&  &  6&  & \\
42\factor& 13:36:52.86& -29:52:50.0& 0.29& \multicolumn{2}{c}{Extended}& 0.51& & 13& 1& 25&  5.9$\times 10^{-16}$& & &  & & &  &  &  5&  &  &  & B3\\
& & & &  &   &   0.01&  &  & & &  2.5$\times 10^{-16}$& & &  & & &  &  &  2&  &  &  & \\
\end{tabular}
\begin{flushleft}
\begin{small}
\factor)~\He PSF or spectroscopic photometry was unavailable and the average slit loss correction factor has been assumed,
\red)~the average E(B--V) value has been assumed,
\extended)~photometry corresponds to a lower limit as the region is extended.  
\end{small}
\end{flushleft}
\end{sidewaystable*}
\begin{sidewaystable*}[h]
{\bf{Table A1.}} (continued).\\
\begin{tabular}{c
@{\hspace{1mm}}c
@{\hspace{2mm}}c
@{\hspace{2mm}}c
@{\hspace{2mm}}c
@{\hspace{1mm}}c
@{\hspace{1mm}}l
@{\hspace{1mm}}c
@{\hspace{1mm}}c
@{\hspace{1mm}}r 
@{\hspace{2mm}}c 
@{\hspace{0.5mm}}c 
@{\hspace{1mm}}r 
@{\hspace{2mm}}c 
@{\hspace{0.5mm}}c 
@{\hspace{1mm}}r 
@{\hspace{2mm}}c 
@{\hspace{0.5mm}}c 
@{\hspace{0mm}}c 
@{\hspace{1.5mm}}c
@{\hspace{2.5mm}}c
@{\hspace{1.5mm}}c
@{\hspace{1.5mm}}c
@{\hspace{1.5mm}}c
@{\hspace{0.5mm}}
}
\noalign{\hrule\smallskip}
Cat & RA & Dec                 & $\rho/\rho_0$ & $m_{4686}$ & $\Delta m$ 
& $E_{\rm B-V}$ & $M_{\rm B}$ & DeV & \multicolumn{3}{c}{C\,{\sc iii} $\lambda$4650/He\,{\sc ii} 4686} & 
\multicolumn{3}{c}{C\,{\sc iii} $\lambda$5696} & 
\multicolumn{3}{c}{
C\,{\sc iv} $\lambda$5808} & \multicolumn{2}{c}{WN} 
& \multicolumn{3}{c}{WC} & MOS\\   
\# & \multicolumn{2}{c}{J2000} &  & mag & mag & mag & mag &
& $W_{\lambda}$ & FWHM & $F_{\lambda}$ 
& $W_{\lambda}$ & FWHM & $F_{\lambda}$ 
& $W_{\lambda}$ & FWHM & $F_{\lambda}$ & 2--5 & 6--10 & 4--6 & 7 &
8--9 & Mask \\
\noalign{\hrule\smallskip}
61& 13:36:57.95& -29:52:55.0& 0.17&  $<$18.6\extended&   0.1&   0.40&  $<$-11.7&  & 4& 30&  8.3$\times 10^{-16}$& 2& 30&  3.0$\times 10^{-16}$& 2& 50&  4.1$\times 10^{-16}$&  &  &  &  2&  & D4\\
& & & &  0.1&   0.1&   0.05&  0.3&  & & &  1.8$\times 10^{-16}$& & &  6.0$\times 10^{-17}$& & &  7.9$\times 10^{-17}$&  &  &  &  1&  & \\
62& 13:36:58.24& -29:48:16.4& 0.53&  19.8&   -0.4&   0.45&  -9.9&  & 20& 50&  7.6$\times 10^{-16}$& 10& 70&  3.6$\times 10^{-16}$& 15& 50&  3.7$\times 10^{-16}$&  &  &  &  &  4& B2\\
& & & &  0.1&   0.1&   0.05&  0.3&  & & &  1.4$\times 10^{-16}$& & &  7.3$\times 10^{-17}$& & &  7.6$\times 10^{-17}$&  &  &  &  &  1& \\
\end{tabular}
\begin{flushleft}
\begin{small}
\factor)~\He PSF or spectroscopic photometry was unavailable and
the average slit loss correction factor has been assumed,
\red)~the average E(B--V) value has been assumed,
\extended)~photometry corresponds to a lower limit as the region is extended.  
\end{small}
\end{flushleft}
\end{sidewaystable*}
\begin{sidewaystable*}[h]
{\bf{Table A1.}} (continued)\\
\begin{tabular}{c
@{\hspace{1mm}}c
@{\hspace{2mm}}c
@{\hspace{2mm}}c
@{\hspace{2mm}}c
@{\hspace{1mm}}c
@{\hspace{1mm}}l
@{\hspace{1mm}}c
@{\hspace{1mm}}c
@{\hspace{1mm}}r 
@{\hspace{2mm}}c 
@{\hspace{0.5mm}}c 
@{\hspace{1mm}}r 
@{\hspace{2mm}}c 
@{\hspace{0.5mm}}c 
@{\hspace{1mm}}r 
@{\hspace{2mm}}c 
@{\hspace{0.5mm}}c 
@{\hspace{0mm}}r 
@{\hspace{1.5mm}}r
@{\hspace{2.5mm}}r
@{\hspace{1.5mm}}r
@{\hspace{1.5mm}}r
@{\hspace{1.5mm}}c
@{\hspace{0.5mm}}
}
\noalign{\hrule\smallskip}
Cat & RA & Dec& $\rho/\rho_0$ & $m_{4686}$ & $\Delta m$ 
& $E_{\rm B-V}$ & $M_{\rm B}$ & DeV & \multicolumn{3}{c}{C\,{\sc iii} $\lambda$4650/He\,{\sc ii} 4686} & 
\multicolumn{3}{c}{C\,{\sc iii} $\lambda$5696} & 
\multicolumn{3}{c}{
C\,{\sc iv} $\lambda$5808} & \multicolumn{2}{c}{WN} 
& \multicolumn{3}{c}{WC} & MOS\\   
\# & \multicolumn{2}{c}{J2000} &  & mag & mag & mag & mag &
& $W_{\lambda}$ & FWHM & $F_{\lambda}$ 
& $W_{\lambda}$ & FWHM & $F_{\lambda}$ 
& $W_{\lambda}$ & FWHM & $F_{\lambda}$ & 2--5 & 6--10 & 4--6 & 7 &
8--9 & Mask \\
\noalign{\hrule\smallskip}
81& 13:37:03.88& -29:54:31.1& 0.37&  21.6&   -1.2&   0.40&  -7.7&  & & No Data&& 130& 40&  3.4$\times 10 ^{-16}$& 35& 30&  9.1$\times 10 ^{-17}$&  &  &  &  &  3& D1\\
& & & &  0.0&   0.1&   0.05&  0.3&  & & &  & & &  1.4$\times 10 ^{-17}$& & &  1.2$\times 10 ^{-17}$&  &  &  &  &  1& \\
82& 13:37:03.97& -29:52:47.5& 0.15&  20.2&   -0.8&   0.24&  -8.6&  & 45& 30&  1.2$\times 10 ^{-15}$& 30& 40&  6.0$\times 10 ^{-16}$& 20& 30&  3.2$\times 10 ^{-16}$&  &  &  &  & 4& A2,C3 \\
& & & &  0.0&   0.0&   0.04&  0.2&  & & &  8.3$\times 10 ^{-17}$& & &  3.3$\times 10 ^{-17}$& & &  2.6$\times 10 ^{-17}$&  &  &  &  &  1& \\
\end{tabular}
\begin{flushleft}
\begin{small}
\factor)~\He PSF or spectroscopic photometry was unavailable and the average slit loss correction factor has been assumed,
\red)~the average E(B--V) value has been assumed,
\extended)~photometry corresponds to a lower limit as the region is extended.  
\end{small}
\end{flushleft}
\end{sidewaystable*}
\begin{sidewaystable*}[h]
{\bf{Table A1.}} (continued).\\
\begin{tabular}{c
@{\hspace{1mm}}c
@{\hspace{2mm}}c
@{\hspace{2mm}}c
@{\hspace{2mm}}c
@{\hspace{1mm}}c
@{\hspace{1mm}}l
@{\hspace{1mm}}c
@{\hspace{1mm}}c
@{\hspace{1mm}}r 
@{\hspace{2mm}}c 
@{\hspace{0.5mm}}c 
@{\hspace{1mm}}r 
@{\hspace{2mm}}c 
@{\hspace{0.5mm}}c 
@{\hspace{1mm}}r 
@{\hspace{2mm}}c 
@{\hspace{0.5mm}}c 
@{\hspace{0mm}}r 
@{\hspace{1.5mm}}r
@{\hspace{2.5mm}}r
@{\hspace{1.5mm}}r
@{\hspace{1.5mm}}r
@{\hspace{1.5mm}}c
@{\hspace{0.5mm}}
}
\noalign{\hrule\smallskip}
Cat & RA & Dec& $\rho/\rho_0$ & $m_{4686}$ & $\Delta m$ 
& $E_{\rm B-V}$ & $M_{\rm B}$ & DeV & \multicolumn{3}{c}{C\,{\sc iii} $\lambda$4650/He\,{\sc ii} 4686} & 
\multicolumn{3}{c}{C\,{\sc iii} $\lambda$5696} & 
\multicolumn{3}{c}{
C\,{\sc iv} $\lambda$5808} & \multicolumn{2}{c}{WN} 
& \multicolumn{3}{c}{WC} & MOS\\   
\# & \multicolumn{2}{c}{J2000} &  & mag & mag & mag & mag &
& $W_{\lambda}$ & FWHM & $F_{\lambda}$ 
& $W_{\lambda}$ & FWHM & $F_{\lambda}$ 
& $W_{\lambda}$ & FWHM & $F_{\lambda}$ & 2--5 & 6--10 & 4--6 & 7 &
8--9 & Mask \\
\noalign{\hrule\smallskip}

101\factor& 13:37:08.25& -29:53:30.5& 0.31& (22.5)& (-0.3)&  0.55& (-8.7)& & 30& 40&  3.3$\times 10 ^{-16}$& & &  & & &  &  11&  &  &  &  & C1\\
& & & &  &   &   0.05&  &  & & &  1.0$\times 10 ^{-16}$& & &  & & &  &  5&  &  &  &  & \\
102& 13:37:08.40& -29:52:54.9& 0.27&  18.1&   -1.0&   0.34&  -11.1&  48& 6& 40&  1.8$\times 10 ^{-15}$& & &  & & &  &  &  9&  &  &  & A4,C3\\
& & & &  0.1&   0.2&   0.01&  0.2&  & & &  3.1$\times 10 ^{-16}$& & &  & & &  &  &  2&  &  &  & \\
103& 13:37:08.53& -29:52:12.0& 0.24&  18.7&   -0.3&   0.78&  -13.2&  49& 6& 40&  1.2$\times 10 ^{-15}$& & &  & & &  &  &  29&  &  &  & A4,C3\\
\end{tabular}
\begin{flushleft}
\begin{small}
\factor)~\He PSF of spectroscopic photometry was unavailable and the average slit loss correction factor has been assumed,
\red)~the average E(B--V) value has been assumed,
\extended)~photometry corresponds to a lower limit as the region is extended.  
\end{small}
\end{flushleft}
\end{sidewaystable*}
\begin{sidewaystable*}[h]
{\bf{Table A1.}} (continued)\\
\begin{tabular}{c
@{\hspace{1mm}}c
@{\hspace{2mm}}c
@{\hspace{2mm}}c
@{\hspace{2mm}}c
@{\hspace{1mm}}c
@{\hspace{1mm}}l
@{\hspace{1mm}}c
@{\hspace{1mm}}c
@{\hspace{1mm}}r 
@{\hspace{2mm}}c 
@{\hspace{0.5mm}}c 
@{\hspace{1mm}}r 
@{\hspace{2mm}}c 
@{\hspace{0.5mm}}c 
@{\hspace{1mm}}r 
@{\hspace{2mm}}c 
@{\hspace{0.5mm}}c 
@{\hspace{0mm}}c 
@{\hspace{1.5mm}}c
@{\hspace{2.5mm}}c
@{\hspace{1.5mm}}c
@{\hspace{1.5mm}}c
@{\hspace{1.5mm}}c
@{\hspace{0.5mm}}
}
\noalign{\hrule\smallskip}
Cat & RA & Dec& $\rho/\rho_0$ & $m_{4686}$ & $\Delta m$ 
& $E_{\rm B-V}$ & $M_{\rm B}$ & DeV & \multicolumn{3}{c}{C\,{\sc iii} $\lambda$4650/He\,{\sc ii} 4686} & 
\multicolumn{3}{c}{C\,{\sc iii} $\lambda$5696} & 
\multicolumn{3}{c}{
C\,{\sc iv} $\lambda$5808} & \multicolumn{2}{c}{WN} 
& \multicolumn{3}{c}{WC} & MOS\\   
\# & \multicolumn{2}{c}{J2000} &  & mag & mag & mag & mag &
& $W_{\lambda}$ & FWHM & $F_{\lambda}$ 
& $W_{\lambda}$ & FWHM & $F_{\lambda}$ 
& $W_{\lambda}$ & FWHM & $F_{\lambda}$ & 2--5 & 6--10 & 4--6 & 7 &
8--9 & Mask \\
\noalign{\hrule\smallskip}
121& 13:37:12.64& -29:49:50.1& 0.50&  21.1& (-1.1)   &   0.46&  -8.2&  & & &  7.1$\times 10 ^{-16}$& & &  & & &  &  17&  &  &  &  & A1\\
& & & &  0.1&   &   0.02&  0.2&  & & &  7.9$\times 10 ^{-17}$& & &  & & &  &  2&  &  &  &  & \\
122& 13:37:12.68& -29:54:47.4& 0.53&  20.8&   -1.0&   0.40&  -8.3&  & 100& 57&  1.4$\times 10 ^{-15}$& & &  & 50& 60&  7.0$\times 10 ^{-16}$&  &  &  3&  &  & C2-3\\
& & & &  0.0&   0.0&   0.01&  0.2&  & & &  1.1$\times 10 ^{-16}$& & &  & & &  1.4$\times 10 ^{-16}$&  &  &  1&  &  & \\
123\factor& 13:37:13.05& -29:51:37.7& 0.39& (21.5)& (0.0)&  0.40& (-9.5)& & 15& 27&  5.7$\times 10 ^{-16}$& 6& 25&  1.5$\times 10 ^{-16}$& & &  &  &  &  &  &  2& A4\\
\end{tabular}
\begin{flushleft}
\begin{small}
\factor)~\He PSF or spectrscopic photometry was unavailable and the average slit loss correction factor has been assumed,
\red)~the average E(B--V) value has been assumed,
\extended)~photometry corresponds to a lower limit as the region is extended.  
\end{small}
\end{flushleft}
\end{sidewaystable*}

\begin{table*}
{\bf{Table A2.}} Catalogue of sources with candidate WR 
signatures in M\,83, either from narrow-band imaging or for which MOS spectroscopy was ambiguous.\\
\begin{center}
\begin{tabular}{l@{\hspace {1.5mm}}l@{\hspace {1.5mm}}l@{\hspace {1.5mm}} c@{\hspace {1.5mm}} c@{\hspace {1.5mm}}c@{\hspace {1.5mm}} c@{\hspace {3mm}}
l@{\hspace {1.5mm}} l@{\hspace {1.5mm}} l @{\hspace {1.5mm}}c @{\hspace {1.5mm}}c@{\hspace {1.5mm}} c@{\hspace 
{1.5mm}} c @{\hspace {1.5mm}}}\hline
Cat&RA&Dec&$\rho / \rho_{0}$&$m_{4686}$& $m_{4686 - 4781}$&Dev&
Cat&RA&Dec&$\rho / \rho_{0}$&$m_{4686}$& $m_{4686-4781}$&Dev\\
&J2000&J2000&&&&&&J2000&J2000&&&&\\
\hline
133&13:36:44.01&--29:51:47.1&0.53&--&--&&178&13:37:01.25&--29:52:52.8&0.13&19.2&--0.2&\\        
134&13:36:44.02&--29:52:26.5&0.54&21.7&--&&179&13:37:01.91&--29:52:13.2&0.05&--&--&\\
135&13:36:44.48&--29:52:10.5&0.52&--&--&&180$^l$&13:37:02.04&--29:55:30.2&0.51&16.9&--0.3&36\\
136&13:36:44.51&--29:53:22.1&0.57&19.2&--0.2&&181&13:37:02.19&--29:57:20.2&0.77&20.6&--0.2&\\
\smallskip
137&13:36:44.55&--29:52:10.0&0.52&--&--&&182&13:37:02.24&--29:55:33.8&0.52&19.9&--0.3&\\
138&13:36:46.05&--29:52:49.7&0.49&19.9&--0.2&&183&13:37:02.64&--29:49:40.8&0.34&18.0&--0.4&39\\
\end{tabular}
\end{center}
${^p)}$ Insufficient S/N for classification, ${^l)}$
insufficient spectral coverage for WN identification, \extended)~photometry corresponds to a lower limit as the region is extended.\\
\end{table*}

\begin{figure*}[h]
\includegraphics[width=18cm,angle=0]{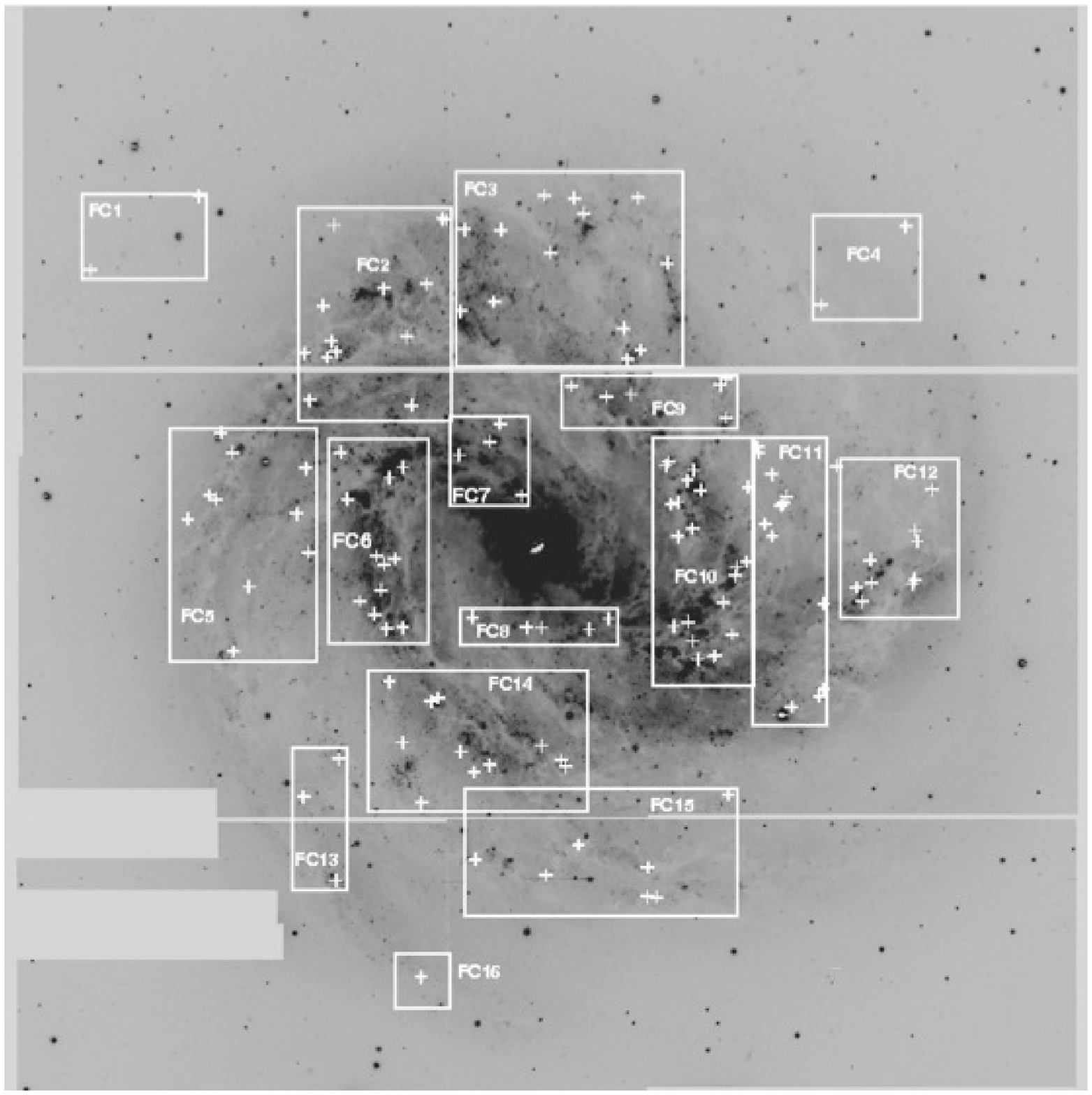}
{\bf Fig. B1:} Master finding chart, indicating location of
confirmed WR sources in M\,83 (+) and corresponding finding chart.
North is up and east to the left on this
$\lambda$4684 narrow-band FORS2 image of M\,83 (12$\times$12 arcmin).
\label{fc1}
\end{figure*}

\clearpage

\begin{figure*}
\includegraphics[width=18cm,angle=0]{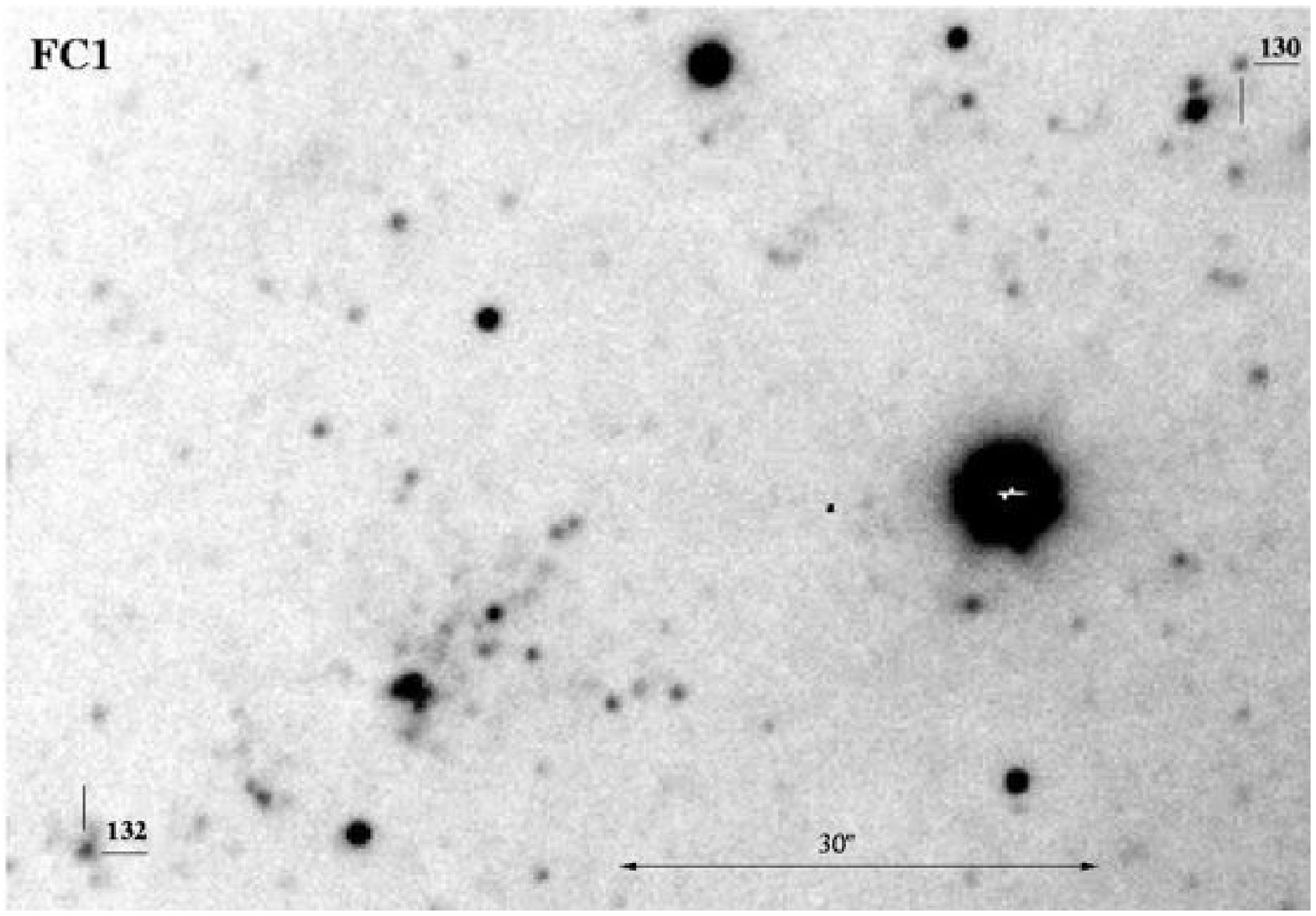}
{\bf Fig. B2:} Finding chart 1. North is up and east to the left on this
$\lambda$4684 narrow-band FORS2 image, with the scale indicated.
\label{fc2}
\end{figure*}

\clearpage

\begin{figure*}
\includegraphics[width=18cm,angle=0]{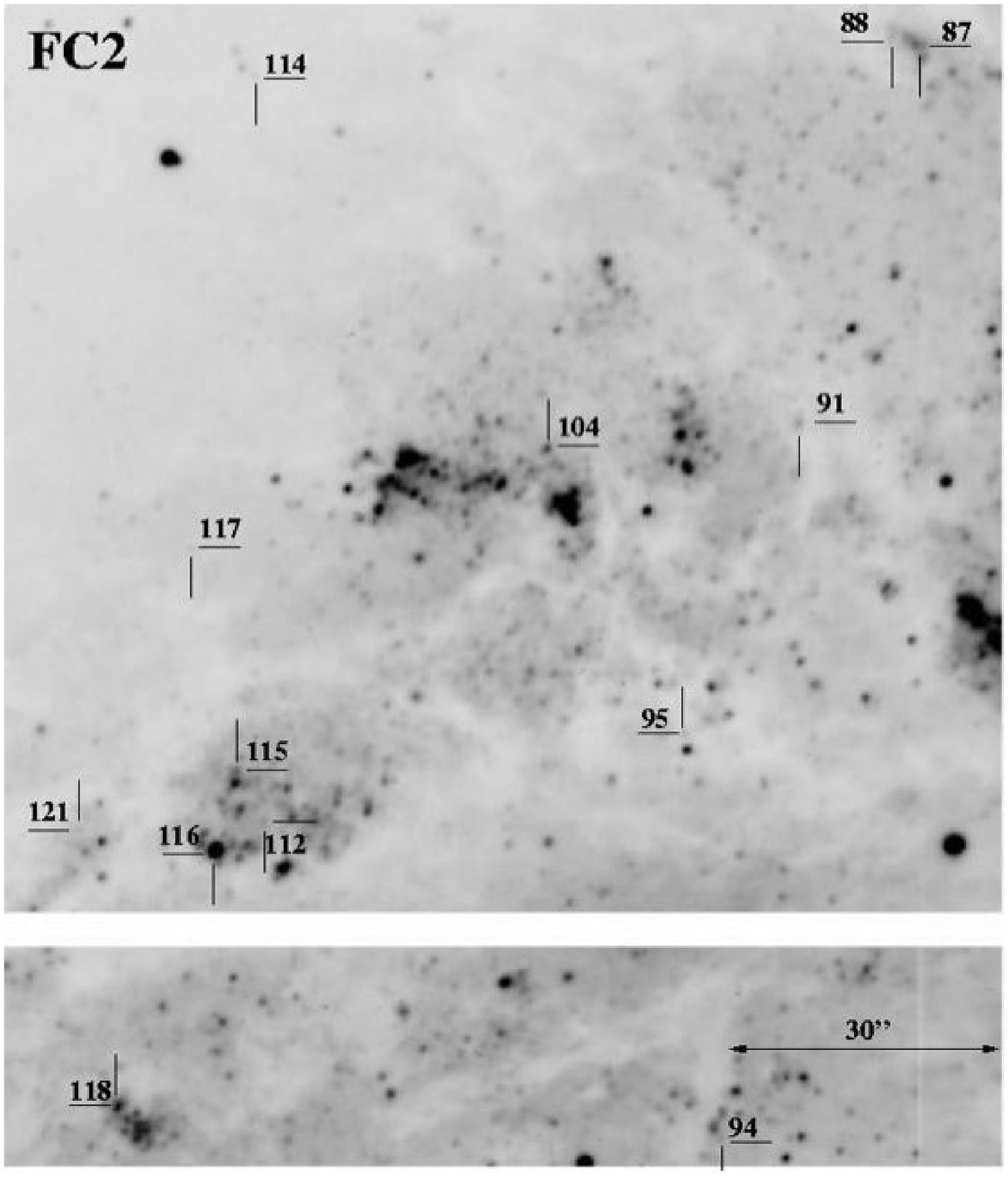}
{\bf Fig. B3:} Finding chart 2. North is up and east to the left on this
$\lambda$4684 narrow-band FORS2 image, with the scale indicated. 
The FORS2
CCD detector gap runs along the bottom of the image.
\label{fc3}
\end{figure*}

\clearpage

\begin{figure*}
\includegraphics[width=18cm,angle=0]{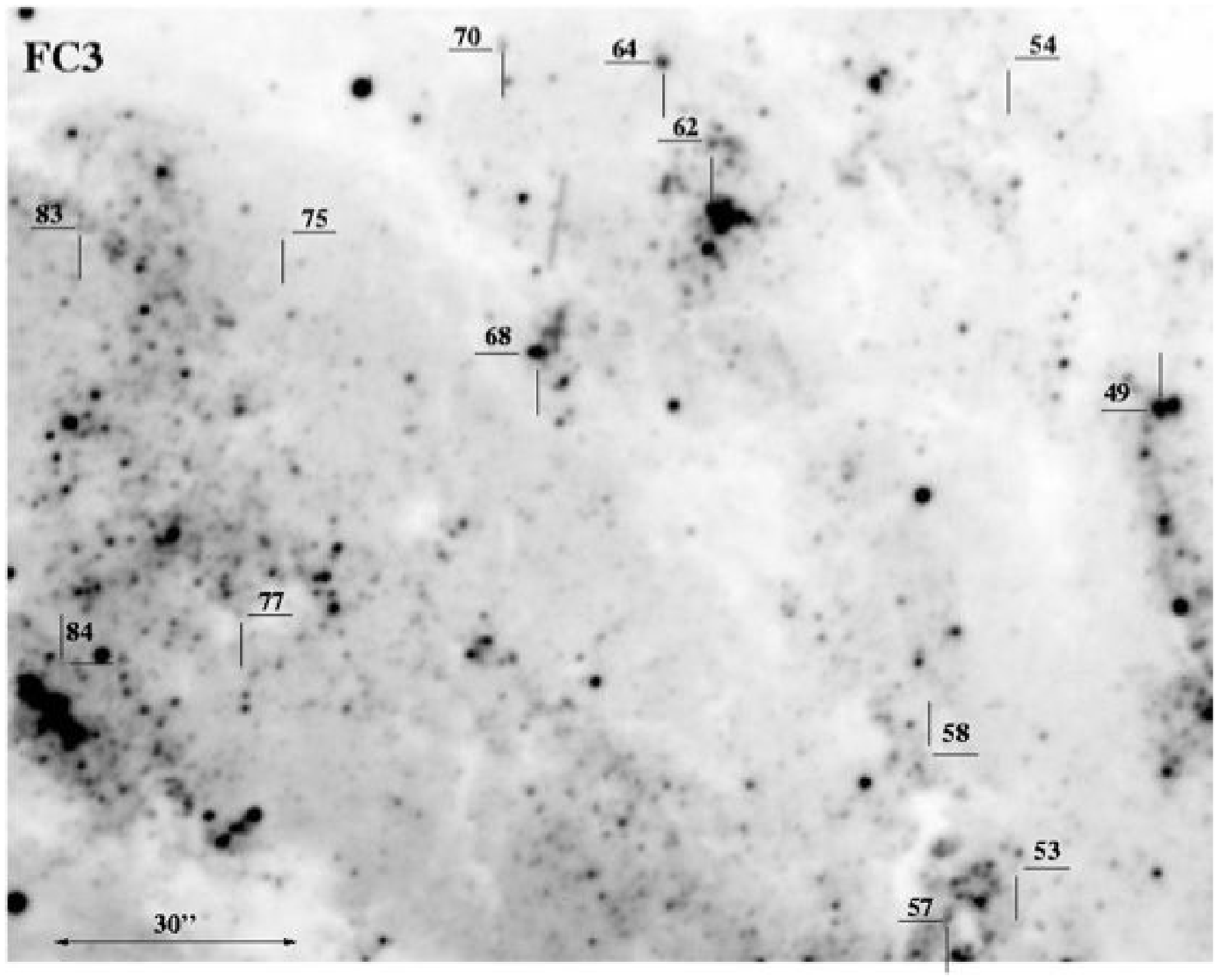}
{\bf Fig. B4:} Finding chart 3. North is up and east to the left on this
$\lambda$4684 narrow-band FORS2 image, with the scale indicated.
\label{fc4}
\end{figure*}

\clearpage
\begin{figure*}
\includegraphics[width=18cm,angle=0]{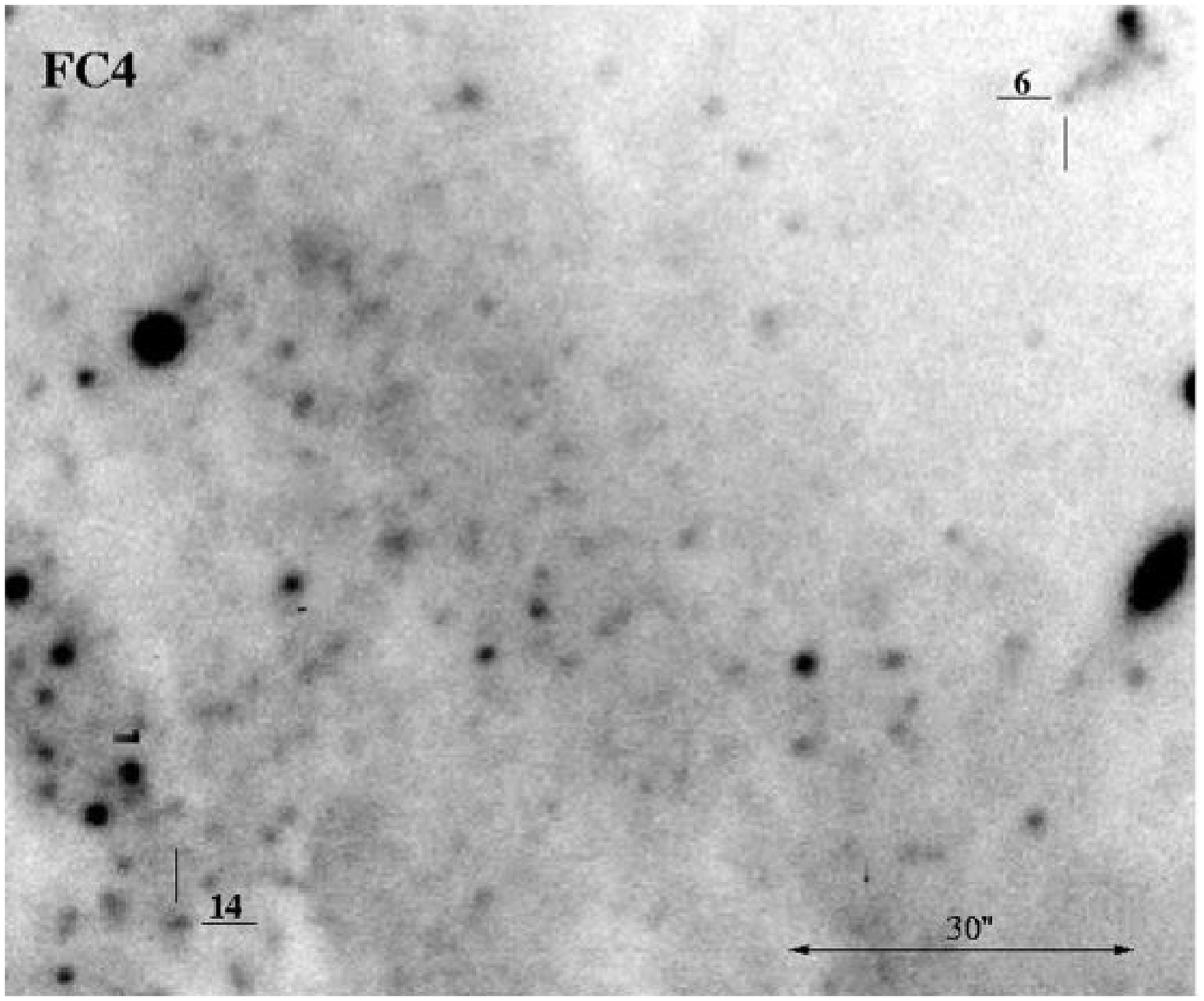}
{\bf Fig. B5:} Finding chart 4. North is up and east to the left on this
$\lambda$4684 narrow-band FORS2 image, with the scale indicated.
\label{fc5}
\end{figure*}

\clearpage

\begin{figure*}
\includegraphics[height=25cm,angle=0]{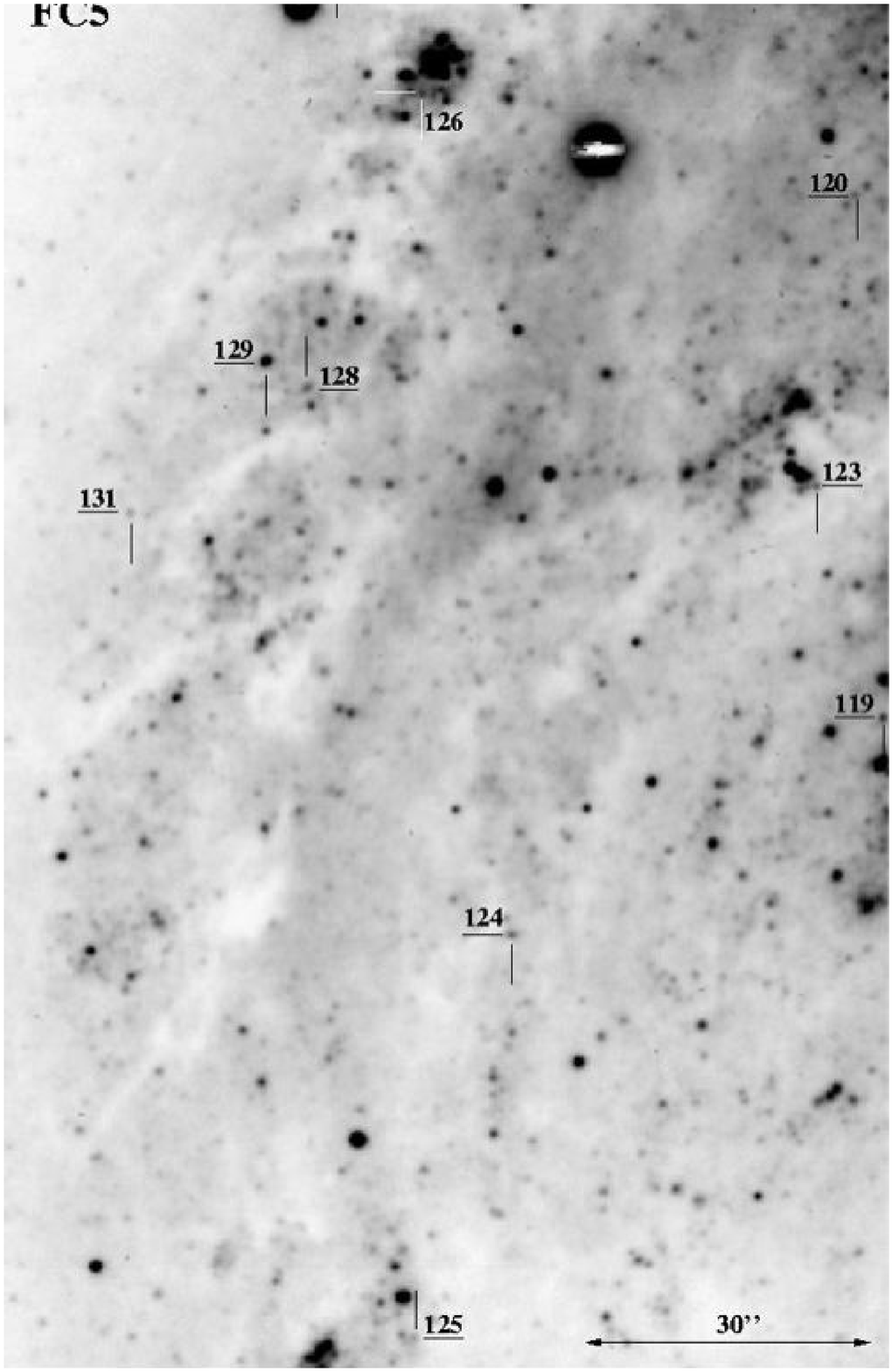}
{\bf Fig. B6:} Finding chart 5. North is up and east to the left on this
$\lambda$4684 narrow-band FORS2 image, with the scale indicated.
\label{fc6}
\end{figure*}

\clearpage

\begin{figure*}
\includegraphics[height=25cm,angle=0]{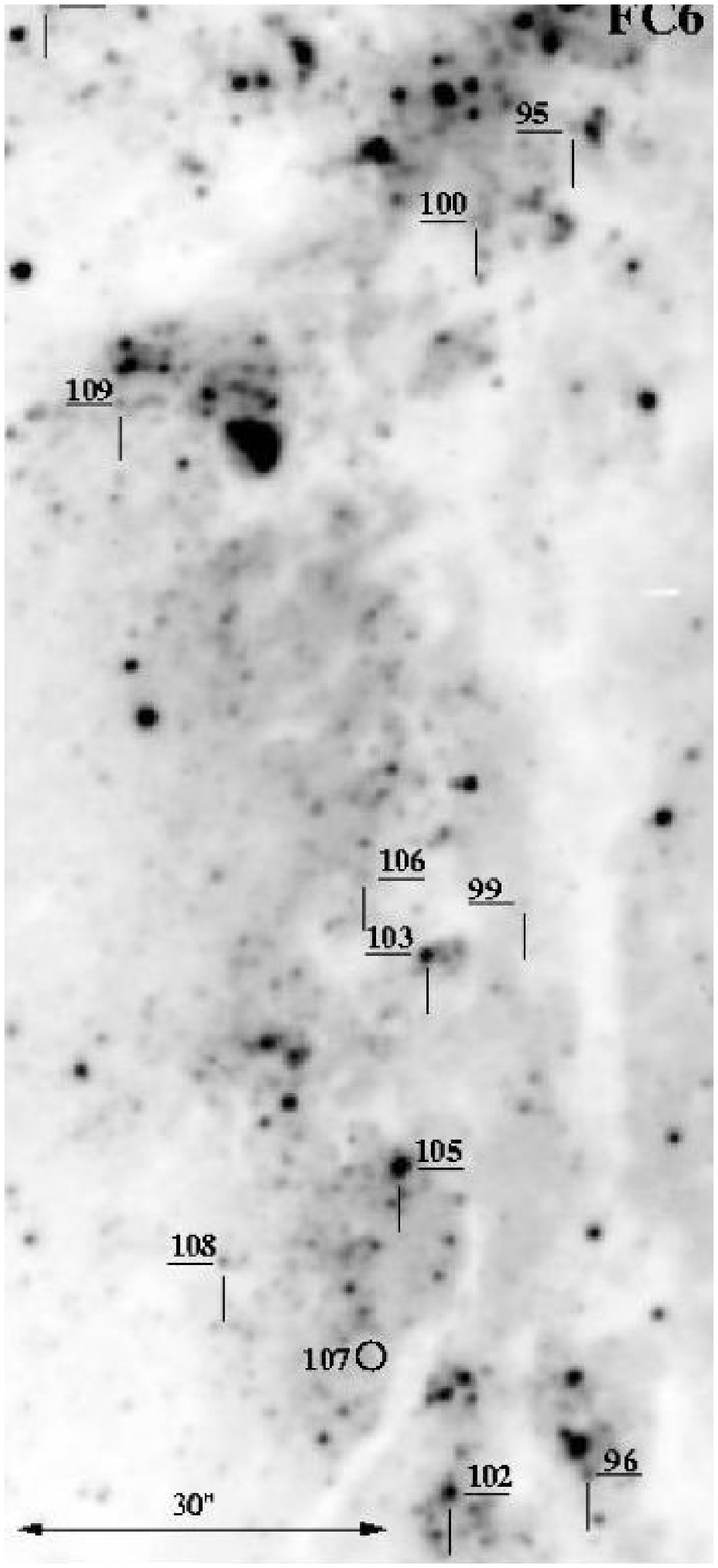}
{\bf Fig. B7:} Finding chart 6. North is up and east to the left on this
$\lambda$4684 narrow-band FORS2 image, with the scale indicated.
\label{fc7}
\end{figure*}

\clearpage

\begin{figure*}
\includegraphics[width=12cm,angle=0]{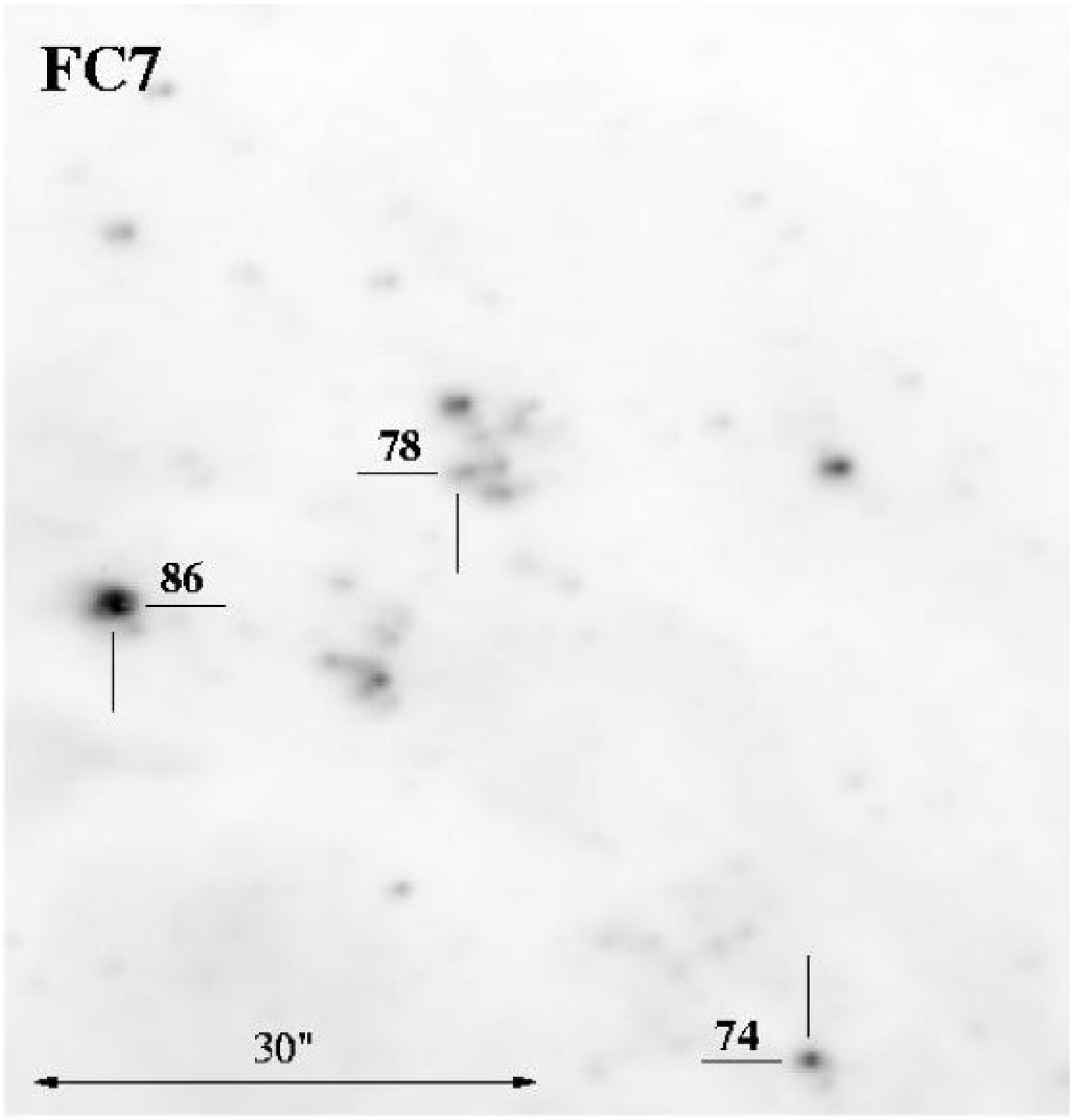}
{\bf Fig. B8:} Finding chart 7. North is up and east to the left on this
$\lambda$4684 narrow-band FORS2 image, with the scale indicated.
\label{fc8}
\end{figure*}

\clearpage

\begin{figure*}
\includegraphics[width=18cm,angle=0]{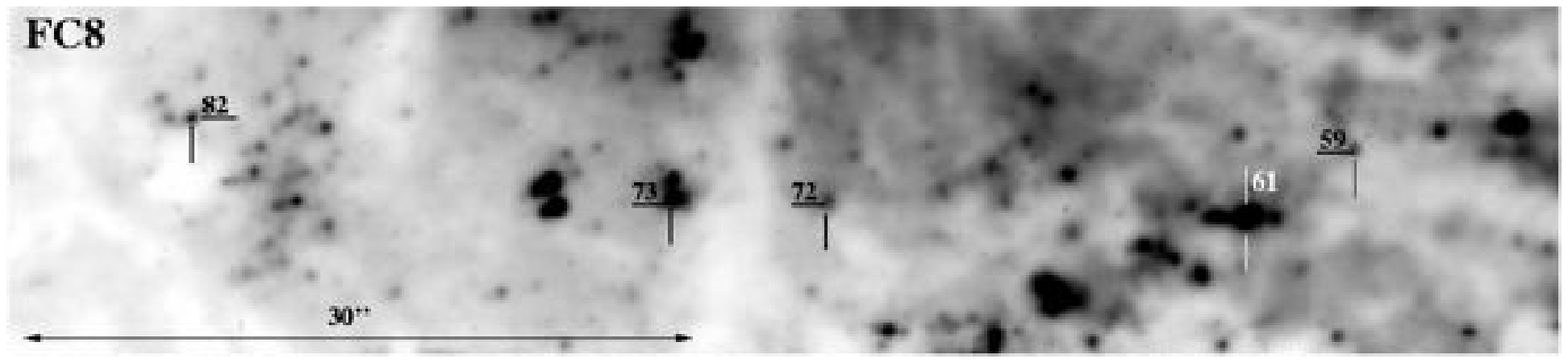}
{\bf Fig. B9:} Finding chart 8. North is up and east to the left on this
$\lambda$4684 narrow-band FORS2 image, with the scale indicated.
\label{fc9}
\end{figure*}

\clearpage

\begin{figure*}
\includegraphics[width=18cm,angle=0]{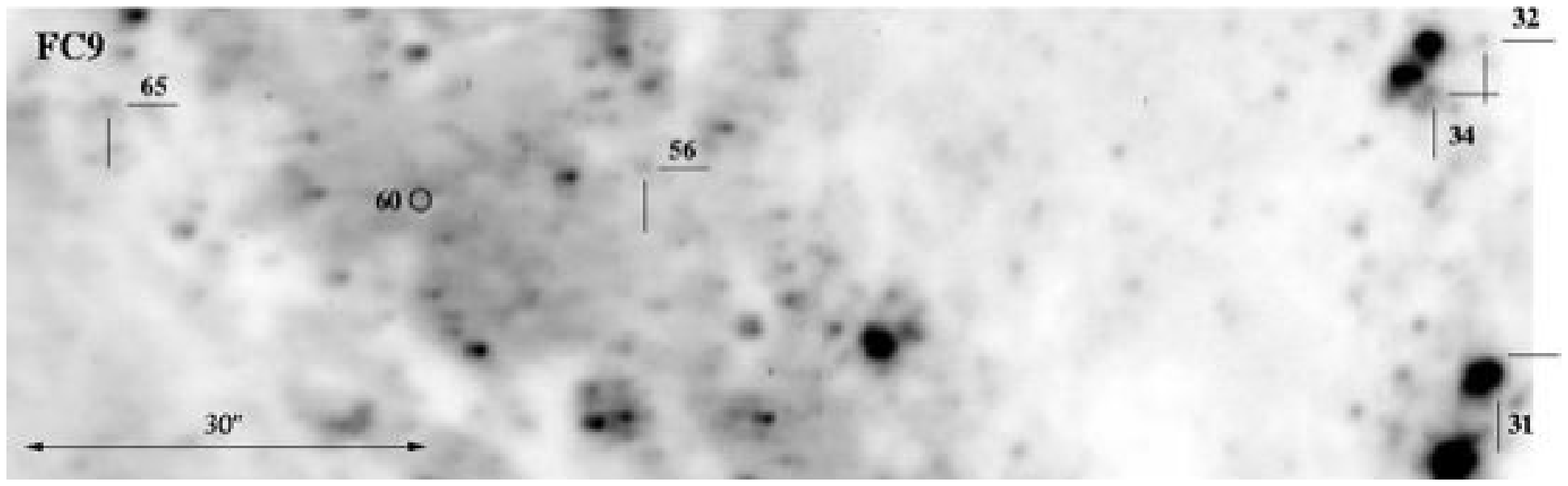}
{\bf Fig. B10:} Finding chart 9. North is up and east to the left on this
$\lambda$4684 narrow-band FORS2 image, with the scale indicated.
\label{fc10}
\end{figure*}

\clearpage

\begin{figure*}
\includegraphics[height=25cm,angle=0]{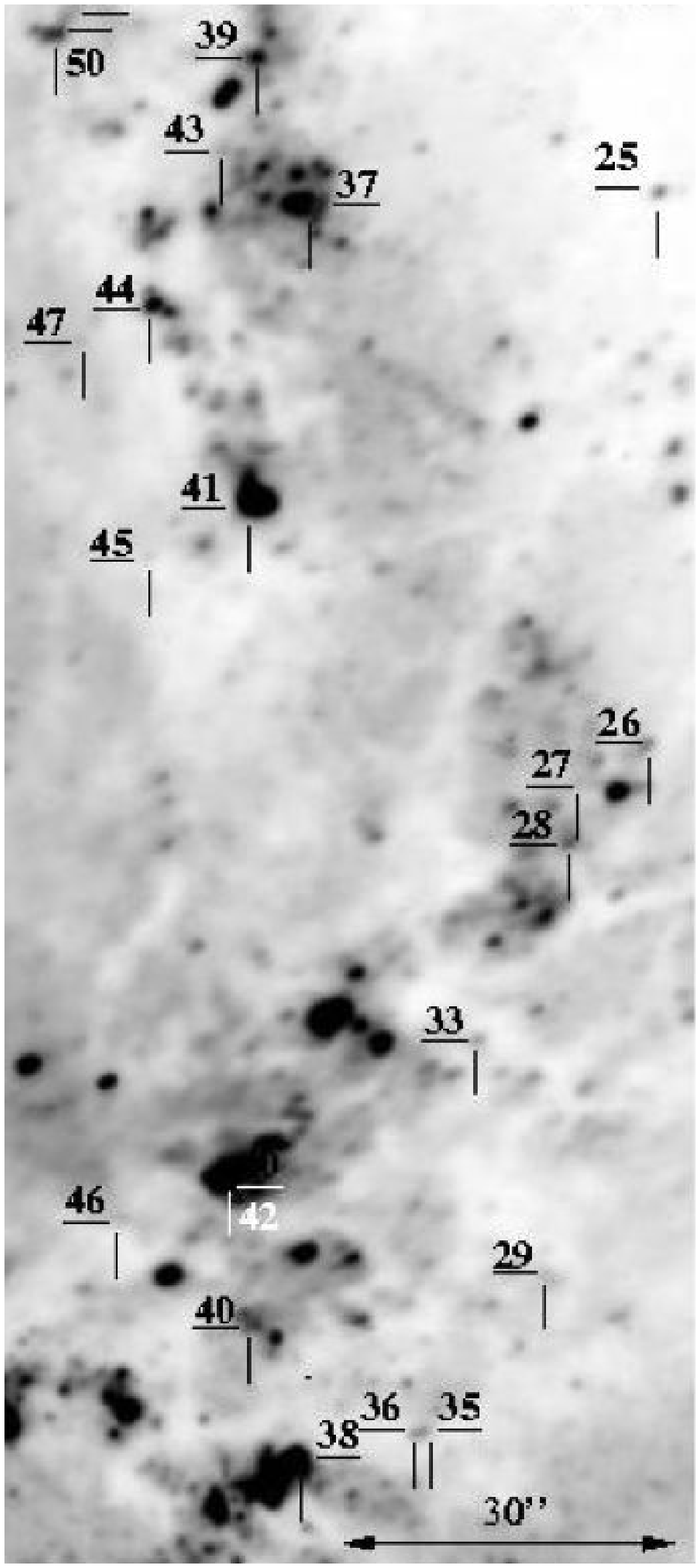}
{\bf Fig. B11:} Finding chart 10. North is up and east to the left on this
$\lambda$4684 narrow-band FORS2 image, with the scale indicated.
\label{fc11}
\end{figure*}

\clearpage

\begin{figure*}
\includegraphics[height=25cm,angle=0]{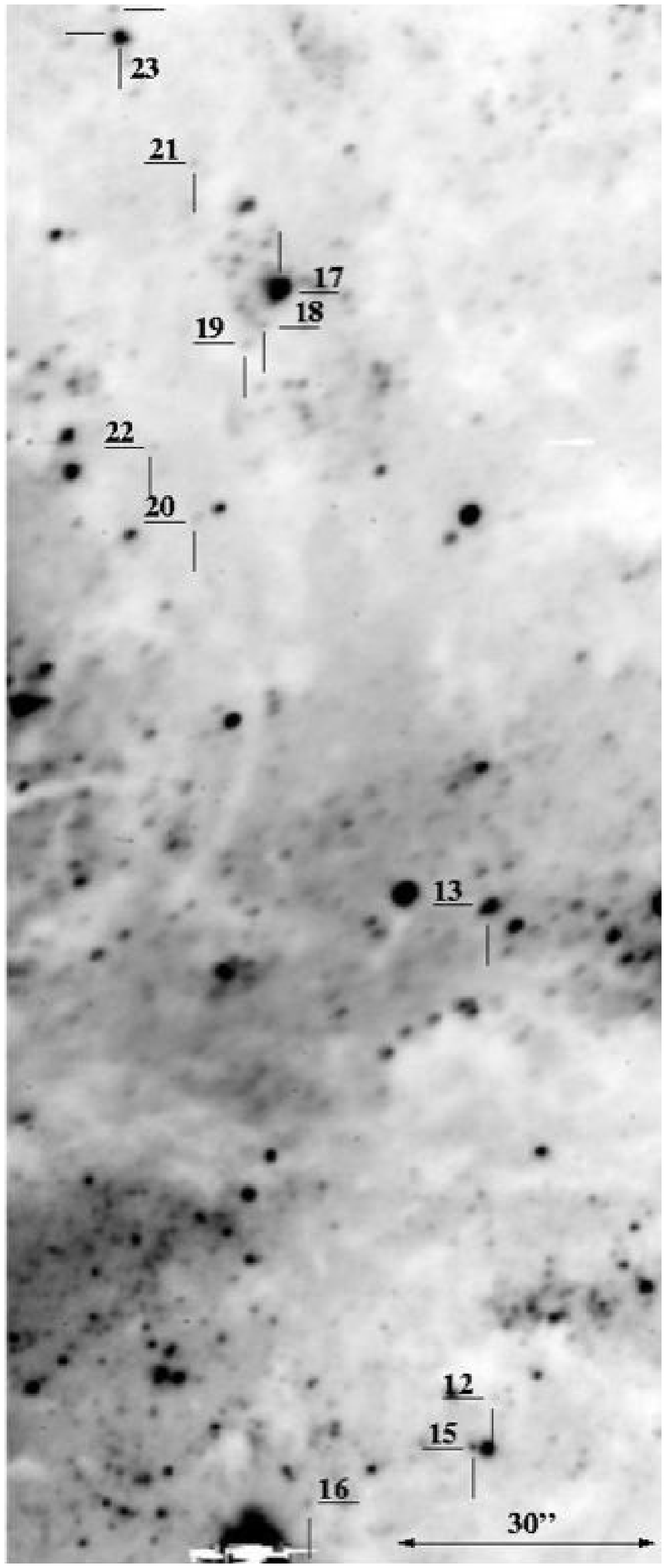}
{\bf Fig. B12:} Finding chart 11. North is up and east to the left on this
$\lambda$4684 narrow-band FORS2 image, with the scale indicated.
\label{fc12}
\end{figure*}

\clearpage

\begin{figure*}
\includegraphics[width=18cm,angle=0]{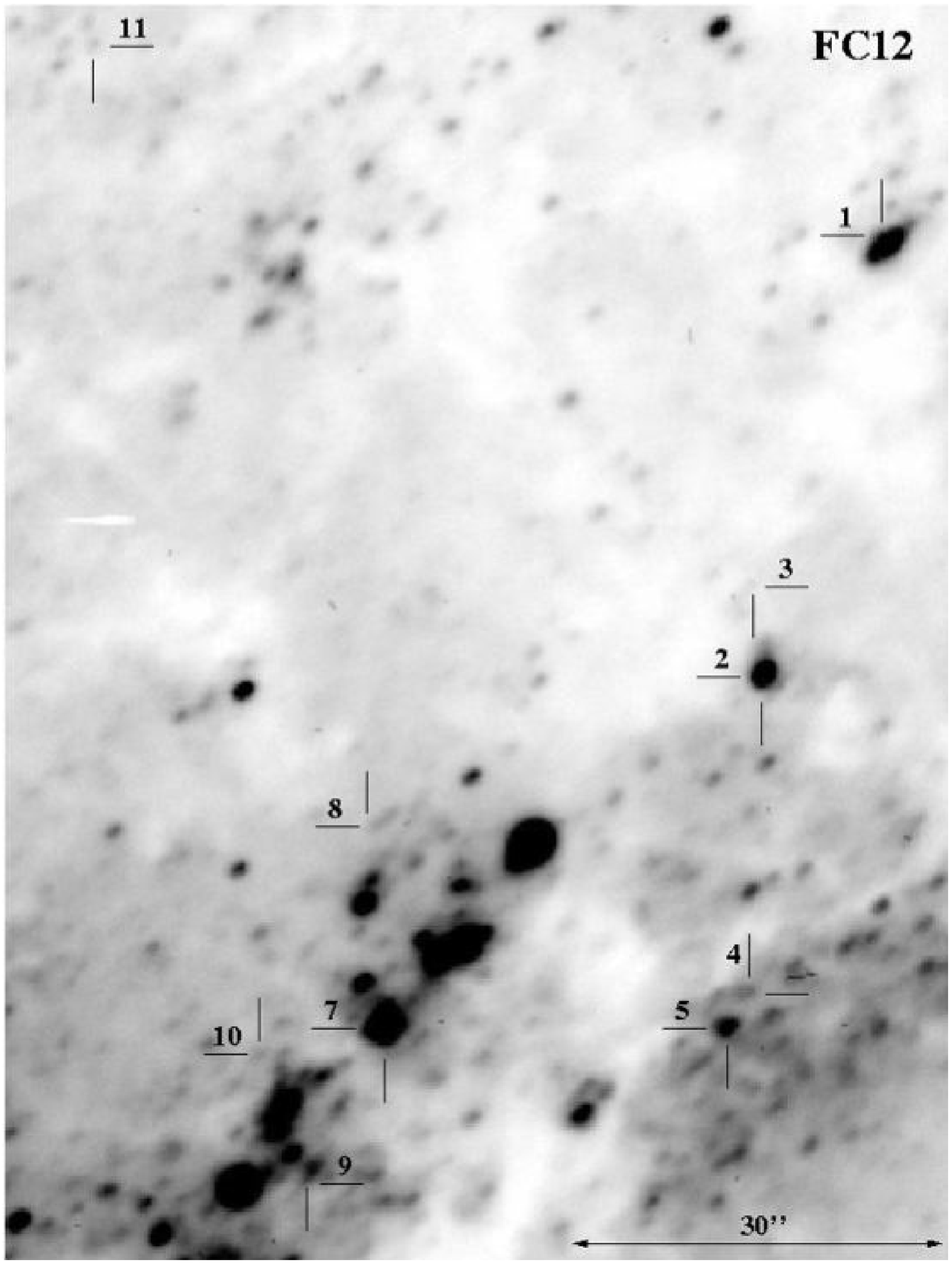}
{\bf Fig. B13:} Finding chart 12. North is up and east to the left on this
$\lambda$4684 narrow-band FORS2 image, with the scale indicated.  The
feature on the west side of the image is a chip defect.
\label{fc13}
\end{figure*}

\clearpage

\begin{figure*}
\includegraphics[height=25cm,angle=0]{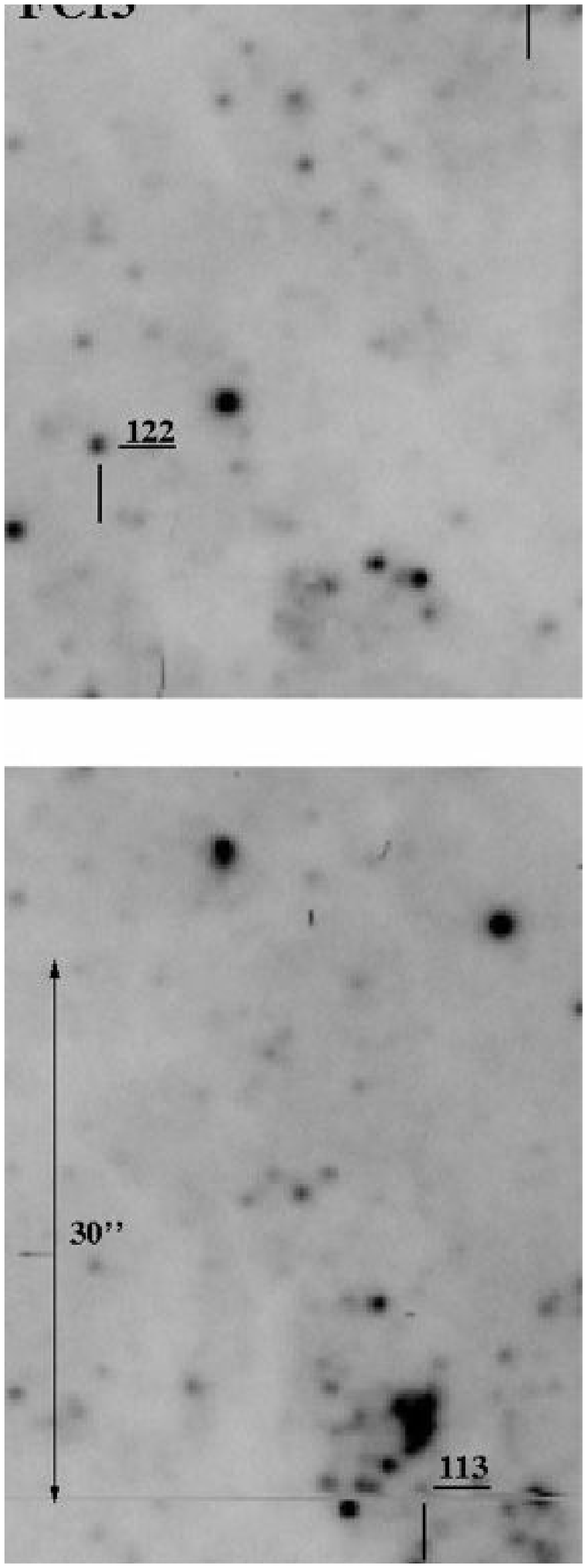}
{\bf Fig. B14:} Finding chart 13. North is up and east to the left on this
$\lambda$4684 narrow-band FORS2 image, with the scale indicated.
The FORS2 CCD detector gap runs along the centre of the image.
\label{fc14}
\end{figure*}

\clearpage

\begin{figure*}
\includegraphics[width=18cm,angle=0]{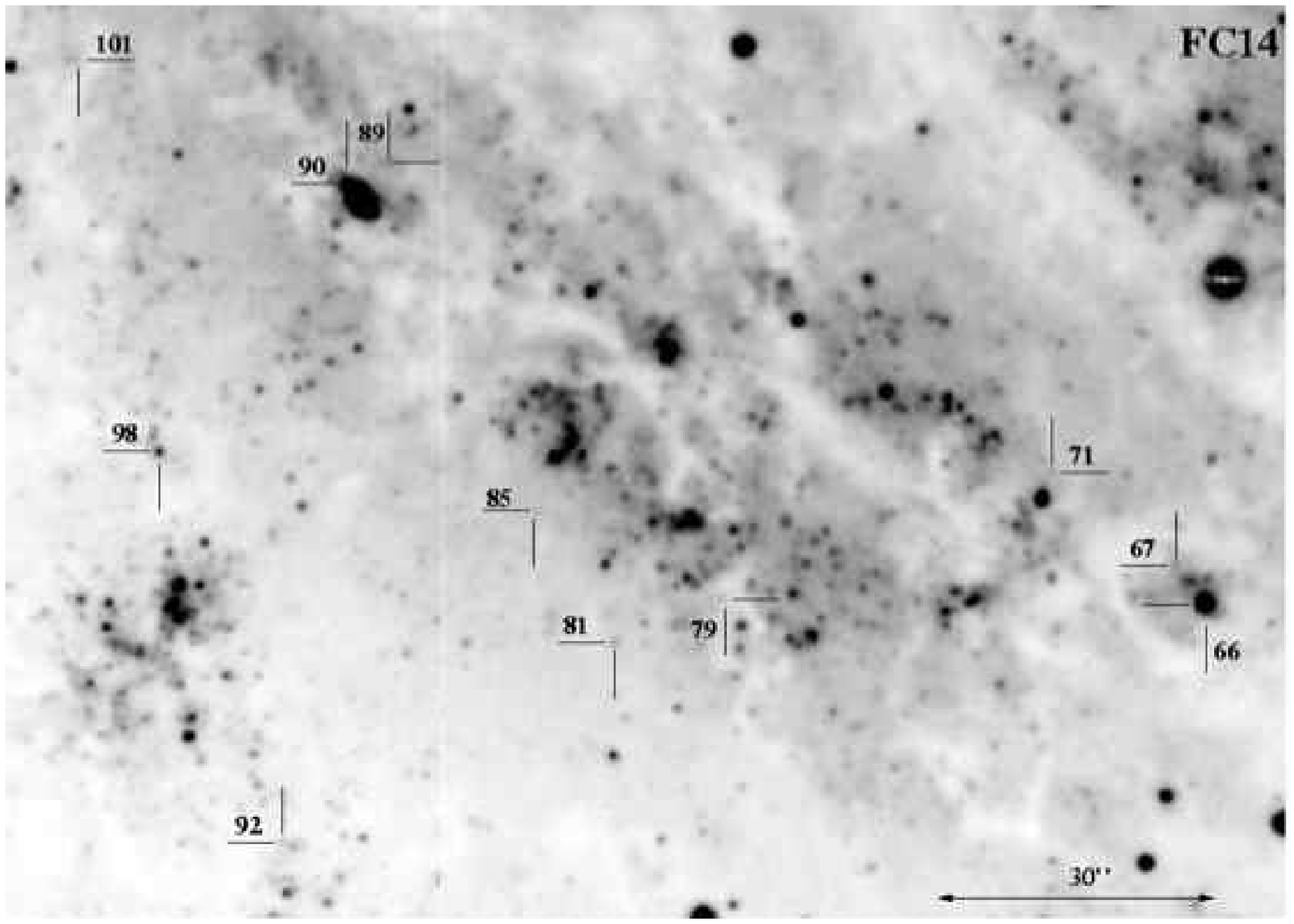}
{\bf Fig. B15:} Finding chart 14. North is up and east to the left on this
$\lambda$4684 narrow-band FORS2 image, with the scale indicated.
\label{fc15}
\end{figure*}

\clearpage

\begin{figure*}
\includegraphics[width=18cm,angle=0]{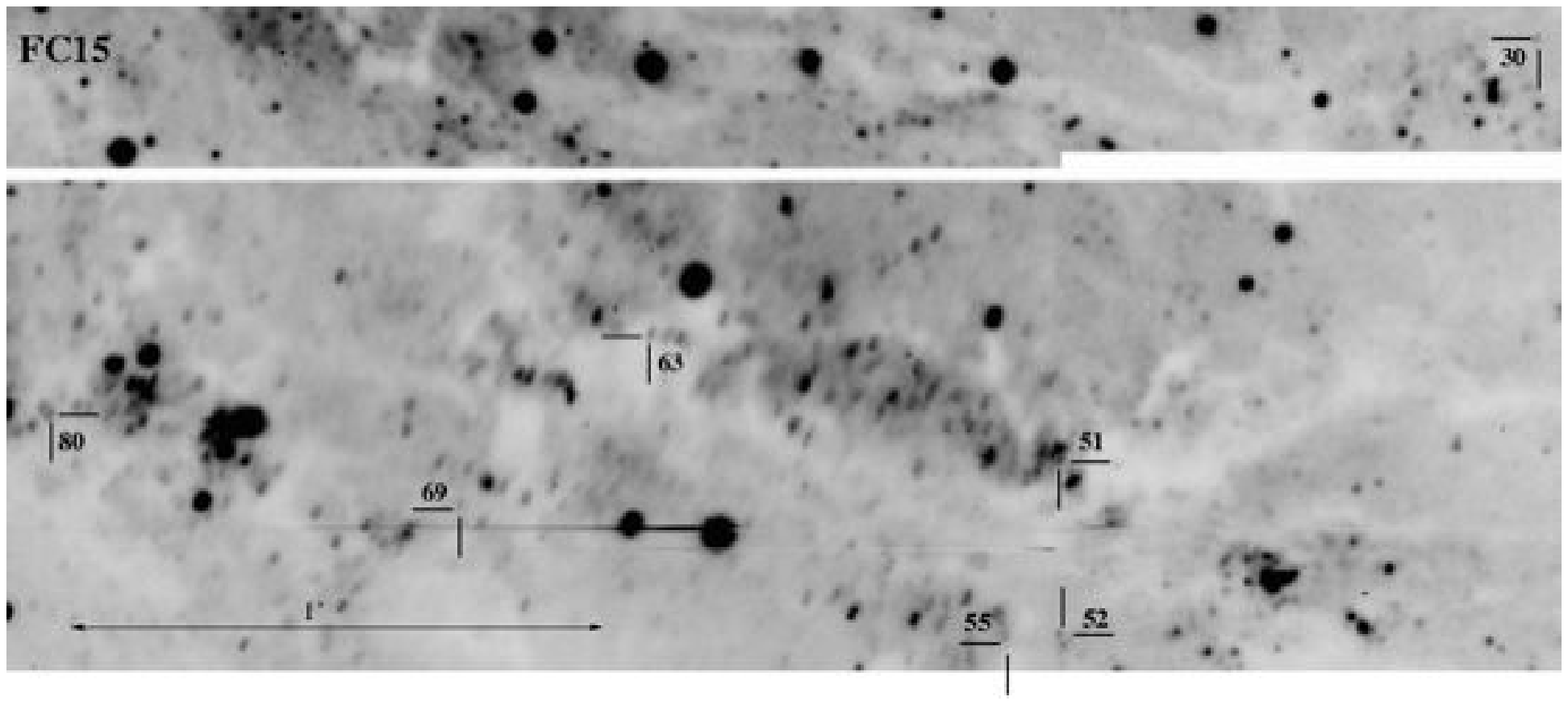}
{\bf Fig. B16:} Finding chart 15. North is up and east to the left on this
$\lambda$4684 narrow-band FORS2 image, with the scale indicated.
The FORS2 CCD detector gap runs along the top of the image.
\label{fc16}
\end{figure*}

\clearpage

\begin{figure*}
\includegraphics[width=9cm,angle=0]{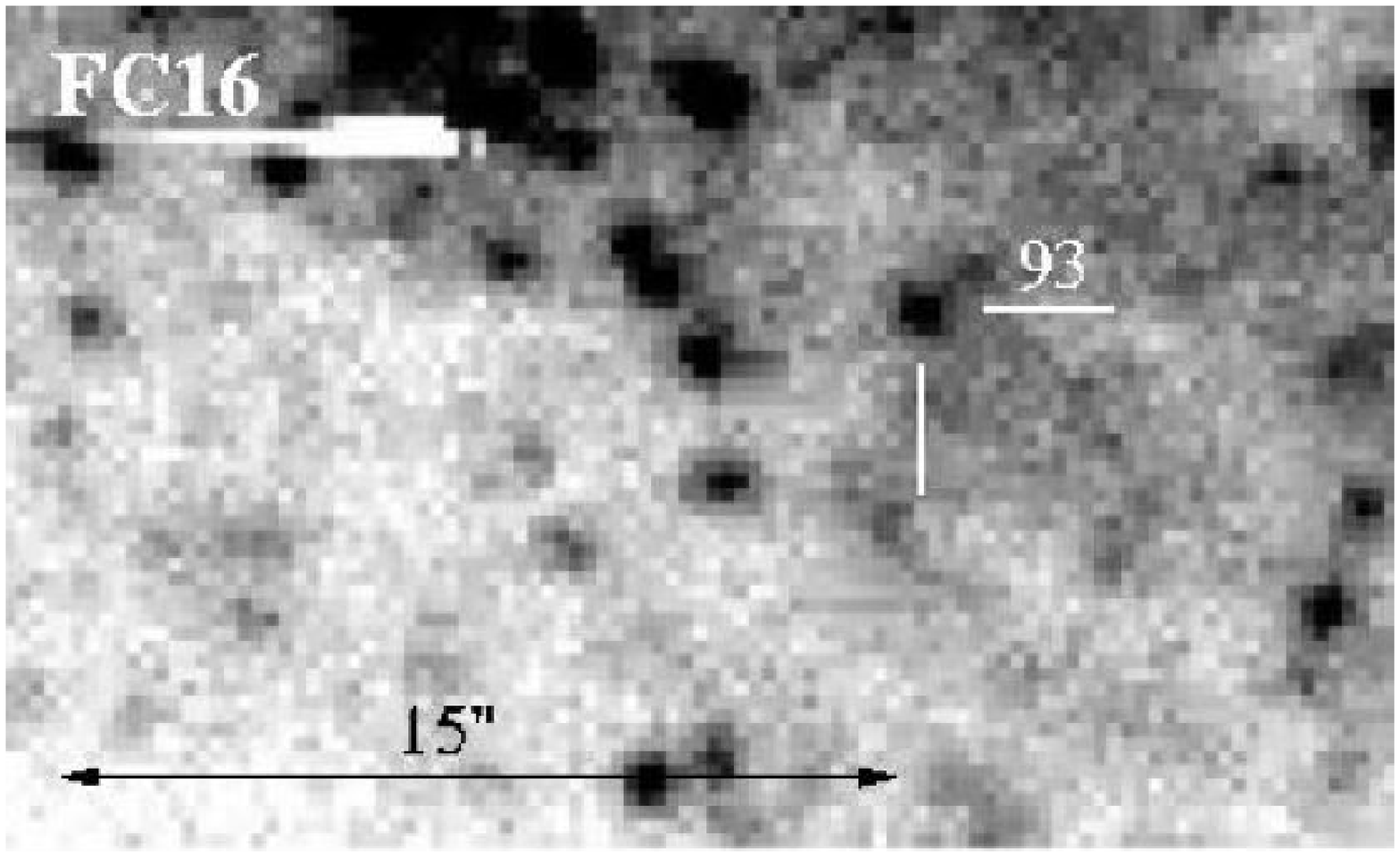}
{\bf Fig. B17:} Finding chart 16. North is up and east to the left on this
$\lambda$4684 narrow-band FORS2 image, with the scale indicated.  The
bright feature in the north west corner of the image is a chip defect.
\label{fc17}
\end{figure*}

\end{document}